\pdfoutput=1
\documentclass[fleqn,usenatbib]{mnras}

\usepackage[T1]{fontenc}

\DeclareRobustCommand{\VAN}[3]{#2}
\let\VANthebibliography\thebibliography
\def\thebibliography{\DeclareRobustCommand{\VAN}[3]{##3}\VANthebibliography}

\usepackage{graphicx}	% Including figure files
\usepackage{amsmath}	% Advanced maths commands
\usepackage{amssymb}	% Extra maths symbols

\usepackage{newtxtext,newtxmath}
\title[The $Z_{\rm abs} - Z_{\rm emiss}$ relation]{A novel approach to investigate chemical inhomogeneities in GRB host galaxies: The $Z_{\rm abs} - Z_{\rm emiss}$ relation}

\author[Metha et al.]{
Benjamin Metha,$^{1, 2}$\thanks{Email: methab@student.unimelb.edu.au}
Alex J. Cameron$^{1, 2, 3}$
Michele Trenti,$^{1,2}$
\\
$^1$School of Physics, The University of Melbourne, VIC 3010, Australia\\
$^2$Australian Research Council Centre of Excellence for All-Sky Astrophysics in 3-Dimensions, Australia \\
$^3$ Department of Physics, University of Oxford, Denys Wilkinson Building, Keble Road,Oxford, OX1 4RH, UK
}

\date{Accepted XXX. Received YYY; in original form ZZZ}

\pubyear{2020}

\begin{document}
\label{firstpage}
\pagerange{\pageref{firstpage}--\pageref{lastpage}}
\maketitle

\begin{abstract}

Models of chemical enrichment and inhomogeneity in high-redshift galaxies are challenging to constrain observationally. In this work, we discuss a novel approach to probe chemical inhomogeneities within long Gamma-Ray Burst (GRB) host galaxies, by comparing the absorption metallicity, $Z_{\rm abs}$, from the GRB afterglow (which probes the environment along the line of sight) with the emission-line metallicity, $Z_{\rm emiss}$, measured via slit spectroscopy.
Using the IllustrisTNG simulation, the theoretical relationship between these metallicity metrics is explored for a range of GRB formation models, varying the GRB progenitor metallicity threshold. For galaxies with fixed $Z_{\rm emiss}$, the median value of $Z_{\rm abs}$ depends strongly on the GRB progenitor threshold metallicity, with $Z_{\rm abs}$ significantly lower than $Z_{\rm emiss}$ for high metallicity hosts. Conversely, at fixed $Z_{\rm abs}$, the median value of $Z_{\rm emiss}$ depends primarily on the metallicity distribution of galaxies in IllustrisTNG and their chemical inhomogeneities, offering a GRB-model-independent way to constrain these processes observationally. 
Currently, only one host galaxy has data for both absorption and emission metallicities (GRB121014A). We re-analyse the emission spectrum and compare the inferred metallicity $Z_{\rm emiss}$ to a recent Bayesian determination of $Z_{\rm abs}$, finding $\log(Z_{\rm emiss}/Z_{\odot}) = \log(Z_{\rm abs}/Z_{\odot}) +0.35^{+ 0.14}_{- 0.25}$, within $\sim 2$ standard deviations of predictions from the IllustrisTNG simulation. Future observations with the \emph{James Webb Space Telescope} will be able to measure $Z_{\rm emiss}$ for 4 other GRB hosts with known $Z_{\rm abs}$ values, using $\sim 2$ hour observations. While small, the sample will provide preliminary constraints on the $Z_{\rm abs}-Z_{\rm emiss}$ relation to test chemical enrichment schemes in cosmological simulations.

\end{abstract}

\begin{keywords}
% One to six of the following: https://static.primary.prod.gcms.the-infra.com/static/site/mnras/document/Updated_keyword_list_Jan_2020_.pdf?node=21e8735032e6494cce9b&version=78171:e4aca8bedd490069ed08
gamma-ray bursts -- ISM:abundances -- software:simulations
\end{keywords}

%%%%%%%%%%%%%%%%%%%%%%%%%%%%%%%%%%%%%%%%%%%%%%%%%%

%%%%%%%%%%%%%%%%% BODY OF PAPER %%%%%%%%%%%%%%%%%%

\section{Introduction}

%It has been established that long gamma-ray bursts (GRBs) act as biased tracers of star formation, with a preference for formation in low-metallicity environments \citep[e.g.][]{Fynbo+03, Modjaz+08, gehrels09, Salvaterra+12, Boissier+13, Perley+13, Vergani+15,Graham&Fruchter2017, Palmerio+2019}. This preference is most naturally explained by a model of GRB formation that includes a metallicity bias, such as the collapsar model, which predicts that stars with metallicities greater than $Z\sim 0.3Z_\odot$ should not be able to produce GRBs \citep{Woosley93, Yoon+06}. 
%Observations of high-metallicity GRB host galaxies may be explained using models of inhomogeneous chemical enrichment, with the site of GRB formation of a lower metallicity than the average metallicity of the host galaxy \citep{Bignone+17, Paper1}. Understanding the details of the interplay between the metallicity bias of GRB progenitors and chemical inhomogeneities in GRB hosts is crucial to understand the origin of GRBs, and could also provide an innovative approach to investigate the assembly history of metals in galaxies.

% %%%%%%%%%%%%%%%%%%%%%%%%% %
% %% Revised text by AJC %% %
% %%%%%%%%%%%%%%%%%%%%%%%%% %

Characterising and understanding the distribution and transport of chemical elements inside galaxies is a critical aspect of galaxy evolution. Successive generations of star formation enrich the interstellar medium (ISM) with metals. Therefore, the spatial distribution of chemical abundances in galaxies is a powerful tracer of the history of gas flows, star formation, accretion, and mergers throughout their assembly (e.g. \citealt{EdmundsGreenhow95, Kewley+10, Torrey+12, Finlator17, Ma+17, Bresolin19, Hemler+20}). 
%A solid theoretical understanding of the interplay between these phenomena is essential both as an input to build realistic subgrid physics models for hydrodynamical cosmological simulations of galaxy formation (e.g., see \citealt{ma2016}), and %for making progress on the outstanding open question of @@BM: Changed because this disagreed with our first sentence. We could keep this, but change our first sentence instead.@@
%understanding whether the connection between massive star formation and long-duration Gamma Ray Bursts is mediated by a metallicity bias (e.g., see \citealt{gehrels09}).
Over the last few decades, our understanding of chemical inhomogeneities in galaxies has advanced dramatically, largely due to the advent of integral field unit (IFU) spectroscopy (see \citealt{MaiolinoMannucci19} for a review). Negative radial metallicity gradients have been widely observed in the low-redshift galaxy population \citep[e.g.][]{Searle71, VilaCostas+92, Berg+13, Berg+20, Ho+15, Belfiore+2017, HP+18} with other studies additionally observing azimuthal variations from these radial metallicity trends \citep{Li+13, Vogt+17, Ho+18, Ho+19, Kreckel+20}. Observational limitations mean that metallicities are more challenging to measure at high-redshift. In the absence of gravitational lensing, spatial resolution is reduced. Furthermore, fainter targets mean that that metallicities are typically derived from fewer emission lines, limiting our ability to control for possible redshift evolution in the ISM conditions of galaxies when constructing an abundance scale (see Appendix~\ref{ap:gas_metallicity} for a more comprehensive discussion). Existing determinations of radial gradients in high-redshift galaxies are generally limited to modest samples of lensed galaxies, or samples of the largest disk galaxies and show substantial amounts of scatter from steep negative gradients to positive gradients \citep{Yuan+11, Swinbank+12, Jones+13, Nicha, Wuyts+16, Carton+18, Wang+19a, Wang+19b, Curti+20b, Gillman+21}. Extending observations to smaller, fainter galaxies remains a challenge. 

Furthermore, it is rare that spatially resolved studies of chemical enrichment in galaxies outside the local Universe have sufficient resolution to characterize local inhomogeneities (rather than just radial variations as a gradient), even though understanding their extent and characteristic scale is important both as an input to build realistic subgrid physics models for hydrodynamical cosmological simulations of galaxy formation (e.g., see \citealt{ma2016}), and for understanding whether the connection between massive star formation and long-duration Gamma-Ray Bursts (GRBs) is mediated by a metallicity bias (e.g., see \citealt{gehrels09, Graham&Fruchter13}).

Absorption spectroscopy is an alternative to emission spectroscopy for probing the metal content and distribution for intrinsically faint high-$z$ objects. This method relies on galaxies being illuminated by a bright background source, such as a quasar, or a GRB \citep{Wolfe+05}. %Such systems are classified by their neutral hydrogen column density: Damped Lyman-$\alpha$ systems (DLAs) are defined to be absorbing systems with $N(HI) > 2\times 10^{20}$cm$^{-2}$; sub-DLAs have $10^{19} \leq N(HI) \leq 10^{20.3}$; and Lyman-limit systems have $3 \times 10^{17} \leq N(HI) \leq 10^{19}$ \citep{Rao+11}. 
Usually, light from background quasars will not pass through the star-forming region of a galaxy, but rather intersect its outskirts \citep{Pettini04, Tumlinson+11, Fumagalli+15}. Furthermore, quasars typically outshine foreground galaxies by several magnitudes, creating an elevated background that prevents the detection of emission lines in the galaxy spectrum, making metallicity measurement via emission-line spectroscopy much more difficult \citep{Ellison+Kewley05}. While hundreds of quasar-illuminated systems have been observed, only a few have measured values of metallicity from emission lines \citep{Noterdaeme+12, Fynbo+13, Wolfe+14}. Because of all this, it is difficult to compare absorption metallicities obtained by quasar DLAs to metallicities determined from the gas phase emissions in the absorption system itself. 

On the other hand, observations of the afterglows of GRBs provide a unique opportunity to probe the metallicities of star-forming regions of high-redshift galaxies without relying on strong emission line diagnostics. The well-established association between long GRBs and Type Ic supernovae shows that GRBs are produced during the collapse of massive stars (e.g. \citealt{GRB=SNe, GRB=SNe2, Cano13}), indicating that GRBs appear in  star-forming regions of galaxies. Because GRB afterglows fade within timescales of a year, 
it is possible to observe GRB host galaxies after the strong ionising background has faded, reducing the difficulty associated with measuring the metallicity of GRB host galaxies via emission-line spectroscopy.

One complication (but also opportunity to test theoretical models of GRB formation) is that GRBs act as biased tracers of star formation, with a preference for formation in low-metallicity environments \citep[e.g.][]{Fynbo+03, Modjaz+08, gehrels09, Salvaterra+12, Boissier+13, Perley+13, Vergani+15,Graham&Fruchter2017, Palmerio+2019}. This preference is most naturally explained by a model of GRB formation that includes a metallicity bias, such as the collapsar model, which predicts that stars with metallicities greater than $Z\sim 0.3Z_\odot$ should not be able to produce GRBs \citep{Woosley93, Yoon+06}. Several high-metallicity GRB host galaxies have been reported, challenging the collapsar framework at face value (e.g. \citealt{Savaglio+12, Heintz+18, Michalowski+18}). These observations may be explained using models of inhomogeneous chemical enrichment, with the site of GRB formation of a lower metallicity than the average metallicity of the host galaxy \citep{Bignone+17, Paper1}. Thus, understanding the details of the interplay between the metallicity bias of GRB progenitors and chemical inhomogeneities in GRB hosts is not only crucial to investigate the origin of GRBs, but it could also provide an innovative approach to investigate the assembly history of metals in high-redshift galaxies.

Currently, a moderate sample ($N=32$) of GRB host galaxies have absorption metallicities measurements 
\citep{Cucchiara+15, Wiseman+17, Bolmer+19}, and proof of concept of emission-line metallicity measurements for one of the hosts in that sample (GRB121024A) was recently presented in \citet{Friis+15}. The host galaxy of GRB121024A at a redshift of $z=2.3$ is relatively bright ($m_{AB} = 22.6$) and massive (estimated stellar mass of $10^{9.9}M_\odot$), and absorption and emission metallicity measurements for this galaxy have previously been reported to be in broad relative agreement ($\log ([O/H])+12 = 8.41^{+0.11}_{-0.12}$ using emission-line spectroscopy and $8.29 \pm 0.12$ using absorption spectroscopy; see \citealt{Arabsalmani+18} and references therein). However, the absorption-based metallicity determination for this galaxy has been recently redetermined by \citet{Bolmer+19} showing a significant difference to the emisison-line metallicity of this galaxy published by \citet{Friis+15}.
A key question that is left unaddressed by the current data and their interpretation is whether the discrepancy in the absorption versus emission metallicities for GRB hosts is generally expected, and to what extent such differences depend on metal production and mixing within high-redshift galaxies as well as on specific aspects of GRB formation models. 

In \citet{Paper1}, we showed for one metallicity biased model of GRB formation that metallicities obtained via observation of the absorption spectrum of the GRB afterglow (hereafter $Z_{\rm abs}$) may be very different to the metallicity obtained using emission-line spectroscopy of the integrated light of the GRB host (hereafter $Z_{\rm emiss}$). In this work, we extend our previous analysis of public datasets from the IllustrisTNG simulation \citep{Illustris1}, exploring how differences between measurements of $Z_{\rm abs}$ and $Z_{\rm emiss}$ for GRB host galaxies depend on redshift, host galaxy properties, and different GRB metallicity bias models. We then explore the feasibility of increasing the sample of GRB hosts with both $Z_{\rm abs}$ and $Z_{\rm emiss}$, taking advantage of the upcoming James Webb Space Telescope (\emph{JWST}). We show that constraining observationally the shape of the $Z_{\rm abs}-Z_{\rm emiss}$ relation for GRB host galaxies could provide information on three different scientific fronts. Firstly, it could be used to test predictions from metallicity biased models of long GRB formation. Secondly, it could shed light on the structure of the interstellar medium (ISM) of high-redshift star forming galaxies in a way that is independent of systematic errors associated with IFU observations. Finally, a well-constrained curve would allow observations of $Z_{\rm abs}$ to be compared directly to $Z_{\rm emiss}$, allowing the gas-phase metallicity of faint GRB hosts to be estimated from observations of the GRB afterglow.

This paper is organised as follows. In Section \ref{sec:models}, our definitions of GRB rate and metallicity in the IllustrisTNG simulation are presented. The theoretical $Z_{\rm abs} - Z_{\rm emiss}$ relation for a variety of GRB models at a range of redshifts is presented in Section \ref{sec:results}. These theoretical relations are compared with observational data in Section \ref{sec:observations}, and opportunities to further constrain this relation with future surveys are discussed. Finally, discussions and conclusions are presented in Sections \ref{sec:discussion} and \ref{sec:conc}, respectively.

\section{Modelling} \label{sec:models}

The IllustrisTNG simulation is a large volume, cosmological, gravo-magnetohydrodynamical simulation run using the moving-mesh code {\sc 
AREPO} \citep{TNG1, TNG2, TNG3, TNG4, TNG5}. %It is the follow-up of the Illustris simulation \citep{Illustris1, Illustris2}, containing updated physical descriptions of galactic winds, AGN feedback, chemical enrichment from stars, metal advection, and black-hole accretion, as well as using improved numerical methods. Unlike Illustris, IllustrisTNG also includes magnetism, overall providing improved descriptions of a large set of observations across redshift and galaxy types. 
As in \citet{Paper1}, this study uses the publically available TNG100-1 simulation. The large $106.5$ cMpc box size helps to decrease cosmic variance \citep{trenti2008}, while the high resolution of this simulation allows variations in ISM conditions in galaxies to be studied down to spatial scales of a few hundred parsecs, over which the metallicity of the ISM has been recently determined to be largely homogeneous \citep{Kreckel+20}.

The internal metallicity distribution of galaxies in the IllustrisTNG simulation arises naturally as a consequence of interactions between models of star formation, the production of metals from these stars, and the redistribution of these metals through advection and galactic winds, as well as interactions with other galaxies including mergers, and secular dynamical processes such as bar-driven radial mixing and spiral arm-driven streaming motions. In IllustrisTNG, star formation events are generated stochastically when the density of a gas cell exceeds a threshold value of $n_H = 0.1$ cm$^{-3}$. Newly-formed stellar populations in TNG100-1 have a total mass of $\sim 10^6 M_\odot$, with an initial mass function following \citet{Chabrier03}, and initial metallicity inherited from the metallicity of the parent gas cell. These stellar populations release metals throughout the asymptotic giant branch phase of their lives as well as when they supernova, with metallicity yields determined from a combination of SPS models, as discussed in \citet{Pillepich+18}. 
Metals are allowed to be transferred between adjacent gas cells passively via advection, or carried to more distant cells via stellar or galactic winds. In the TNG simulation, characteristic outflows velocities from star forming regions decrease with increasing metallicity -- see \citet{Pillepich+18} for further details. 

\subsection{Modelling local and global metallicities}

The metallicity of each gas cell in the IllustrisTNG simulation is defined as the mass ratio of all elements heavier than helium to the total mass of the gas cell.

\begin{equation}\label{eq:zcell}
Z_{\rm cell} := M_{Z}/M_{\rm cell},
\end{equation}

Observationally, at $z \gtrsim 2$, the internal metallicity profile of galaxies can rarely be resolved \citep[e.g.][]{Curti+20b}. Instead, global metallicity measurements of galaxies are far more common. In this model, we assume that high redshift GRB host galaxies may have their metallicities measured in one of two ways: either by emission-line spectroscopy, or absorption spectroscopy of the GRB afterglow.

% AJC edit:
Emission line methods for determining the gas metallicity rely on measurements of gas that is photoionised by bright, young stars in the galaxy. Because the UV brightness of galaxies locally traces star formation, metallicities obtained via emission line spectroscopy predominately trace regions of star formation within galaxies.
Therefore, to reflect this bias, $Z_{\rm emiss}$ is defined to be the star formation rate weighted average metallicity of the simulated galaxy.

\begin{equation} \label{eq:z_emiss}
    Z_{\rm emiss} := \frac{\sum_{\text{all gas cells}} Z_{\rm cell} \times {\rm SFR}_{\rm cell}}{\sum_{\text{all gas cells}}{\rm SFR}_{\rm cell}}
\end{equation}

This definition has no spatial cutoff, and assumes that the angular extent of these galaxies is smaller than the PSF of the observing telescope -- an assumption that is reasonable for high-redshift GRB hosts. Similar approaches can be seen in the literature (e.g. \citealt{TNG_MZR}, \citealt{Yates+20}).

On the other hand, absorption spectroscopy is independent of the SFR. Instead, $Z_{\rm abs}$ yields an unbiased measurement of the mass fraction of metals along the line-of-sight (LOS) between the observer and the GRB. 

\begin{equation}
    \label{eq:Z_DLA}
     Z_{\rm abs} := \frac{\sum_{\text{gas cells along LOS}} Z_{\rm cell} \times \rho_{\rm cell} \times l_{\rm cell}}{\sum_{\text{gas cells along LOS}} \rho_{\rm cell} \times l_{\rm cell}},
\end{equation}
%%%%%%%%%%%
where $l_{cell}$ is the length of the line-of-sight passing through that gas cell, and $\rho_{cell}$ is the density of each gas cell. Following \citet{Vreeswijk+2012}, we assume that GRBs will ionise all material that is more than 100pc from the site of the burst. For this reason, the start of the line-of-sight is set to be 100pc away from the site of the burst, and the end of the line-of-sight is chosen to be at the location of the closest gas cell to the observer.

Note that for galaxies with homogeneous metallicities throughout their ISM, $Z_{\rm abs}$ and $Z_{\rm emiss}$ from Equations~\ref{eq:z_emiss} and \ref{eq:Z_DLA} will yield the same result. Any variance between these two different metallicity measurements can therefore be attributed to chemical inhomogeneities in GRB hosts.

Throughout this paper, metallicities will be reported relative to the metallicity of the Sun.  Following \citet{Asplund09}, we adopt a solar metallicity value of $Z_\odot = 0.0134$.\footnote{Note that this is slightly different to the value of $Z_\odot = 0.0127$ adopted by the TNG collaboration.}

\subsection{Modelling GRB formation}

For each galaxy in this simulation, the rate of GRBs originating from that galaxy is computed in postprocessing. To account for the chemical inhomogeneities present in these galaxies, the rate of GRB production is computed separately for each gas cell in each galaxy. Because GRBs are produced during the collapse of some massive stars, the rate of GRBs originating from a gas cell should be proportional to the star formation rate of that cell \citep{Fruchter+06}, multiplied by some model-dependant metallicity bias function.\footnote{Other relevant factors, such as the fraction of tight binaries formed in each population and the angular momentum distribution of newly formed stars, are not modelled in the IllustrisTNG simulation, and so are assumed to be the same for all gas cells.}

\begin{equation} \label{eq:cell_wise_rate}
    \rho^{(\rm GRB)} = \sum_{\text{all gas cells}} \kappa(Z_{\rm cell})\times \rm{SFR}_{\rm cell}.
\end{equation}

In this study, $\kappa(Z)$ is assumed to take the form of a simple cutoff function. While other more complex forms for the metallicity bias function have been discussed \citep[e.g.][]{Yoon+06, trenti2015, Chrimes+20}, to first approximation these can all be approximated by (a combination of) cutoff functions with appropriate choices for $Z_{\rm max}$.

\begin{equation} \label{eq:cutoff_bias}
    \kappa(Z)=\begin{cases}
      \kappa_{0}, & \text{if}\ Z < Z_{\rm max} \\
      0, & \text{otherwise}
    \end{cases}
\end{equation}

For each model, the constant scaling factor $\kappa_{0}$ is fixed to best match the rate of GRBs as a function of redshift as determined by \citet{SHOALS1}. In this study, because we are selecting the same number of GRB host galaxies at each redshift from each model, our results do not depend on $\kappa_0$. The value of $Z_{\rm max}$ is varied between $0.01Z_\odot$ and $1.0Z_\odot$ to explore the predicted relation between $Z_{\rm abs}$ and $Z_{\rm emiss}$ for different GRB formation models.

\subsection{On the stochastic nature of $Z_{\rm abs}$}

\begin{figure*}
	\includegraphics[width=0.92\textwidth]{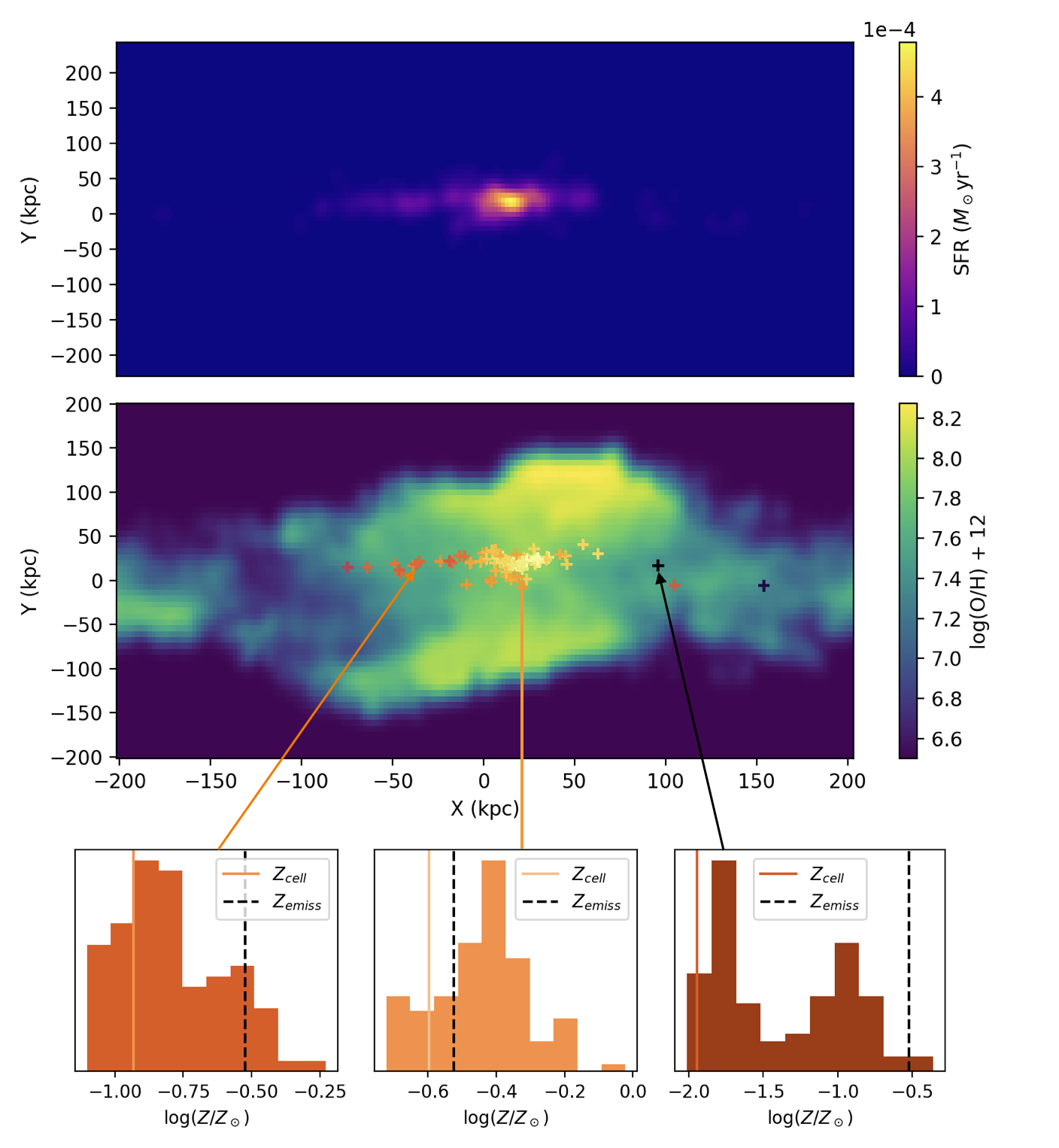}
    \caption{Exploring variations of $Z_{\rm abs}$ within Subhalo 21310 of Snapshot 30 of the TNG100-1 simulation (SFR$=32.86\rm M_\odot$/yr, $\rm M_*=1.29\times 10^{11}\rm M_\odot$, $Z_{\rm emiss}=0.28Z_\odot$). \emph{Top Panel:} SFR profile of this galaxy, used to compute the probability of a GRB originating from each gas cell. Note that the plotted SFR values reflect the average SFR of gas cells within each spatial bin. \emph{Middle panel:} Metallicity profile for this galaxy, with the median value of $Z_{\rm abs}$ for 100 gas cells displayed. From this plot, we can see that GRBs originating from more metal-rich regions of a galaxy produce higher $Z_{\rm abs}$ values along their lines of sight. \emph{Bottom panel:} Histograms showing possible values of $Z_{\rm abs}$ from within the same gas cell, using 100 different orientations. Overplotted on these histograms is the metallicity of the gas cell that hosts the GRB ($Z_{\rm cell}$, solid coloured line), and the overall metallicity of the host galaxy ($Z_{\rm emiss}$, dashed black line).}
    \label{fig:stochastic_Z_abs}
\end{figure*}

Based on Equation \ref{eq:Z_DLA}, the same GRB host galaxy can have vastly different measured values of $Z_{\rm abs}$, depending on (i) the location of the GRB within the host galaxy, and (ii) the orientation of the burst relative to an observer. To illustrate the dependence of $Z_{\rm abs}$ on these effects, the galaxy most likely to be a GRB host in Snapshot 30 of the TNG100-1 simulation was selected, assuming a cutoff metallicity of $0.35 Z_\odot$. This galaxy (Subhalo 21310 in the IllustrisTNG {\sc Subfind} catalog) has a star formation rate of $32.86\rm M_\odot$/yr, a stellar mass of $\rm M_*=1.29\times 10^{11}\rm M_\odot$, and $Z_{\rm emiss}=0.28Z_\odot$. For this galaxy, 100 different gas cells were randomly selected to be the sites of the GRB, with the likelihood of each cell being chosen weighted by its likelihood of producing a GRB progenitor, assuming $Z_{\rm max} = 0.35 Z_\odot$. For each of these locations, 100 different directions were randomly chosen to be the orientations of the lines of sight.

In Figure \ref{fig:stochastic_Z_abs} we show the results of this investigation. The top two panels of this figure show the edge-on projected distributions of $Z$ and SFR for the galaxy; the two properties used to compute the rate of GRB formation for each cell. Plotted over the metallicity map for this galaxy is the locations of the 100 sites from which GRBs were chosen to originate. Points are colour-coded to indicate the median values of $Z_{\rm abs}$ obtained from the 100 LOS trials for each cell. From this figure, it can be seen that when the origin of the GRB LOS is closer to the metal-rich centres of galaxies, $Z_{\rm abs}$ is likely to be higher. Below this metallicity profile, histograms showing the full range of $Z_{\rm abs}$ obtained from all 100 orientations are shown for a selection of gas cells with different metallicities at different locations in this galaxy. Depending on the orientation of the LOS, $Z_{\rm abs}$ can vary by about 1 order of magnitude, even for GRBs originating from the same location within the same galaxy. Furthermore, the range of possible values that $Z_{\rm abs}$ can take increases as the difference between the local GRB host gas cell metallicity $Z_{\rm cell}$ and the global SFR-weighted mean metallicity $Z_{\rm emiss}$ increases.

Due to the wide variance of possible values of $Z_{\rm abs}$ even within the same GRB host galaxy, we caution against using measured values of $Z_{\rm abs}$ to infer global metallicities, $Z_{\rm emiss}$, of individual host galaxies. However, with a large enough sample of galaxies, the median trend between $Z_{\rm abs}$ and $Z_{\rm emiss}$ can be used to shed light on the threshold metallicity for GRB formation, and reveal the presence of chemical inhomogeneities within GRB host galaxies at a wide range of redshifts. We undertake such an analysis using simulated populations of GRB host galaxies from the IllustrisTNG simulation in the next section. 

\section{Results} \label{sec:results}

Four snapshots of the TNG simulation were downloaded at a range of redshifts $(z \in \{1.74, 2.32, 3.0, 4.0 \} )$. Following \citet{Paper1}, all galaxies in each snapshot were identified by using the publicly available dark-matter halo catalogues generated through a friends-of-friends algorithm \citep{FOF}, and searching for collections of subhalos with overlapping half-star radii (the radius enclosing half of the stellar mass). This definition ensures that galaxies in the simulation are photometrically inseparable collections of baryons, and not simply high-density disk instabilities inside other, larger galaxies \citep{TNG2}, or gravitationally distinct subhalos that cannot be separated by observation.

\begin{figure*}
	\includegraphics[width=0.49\textwidth]{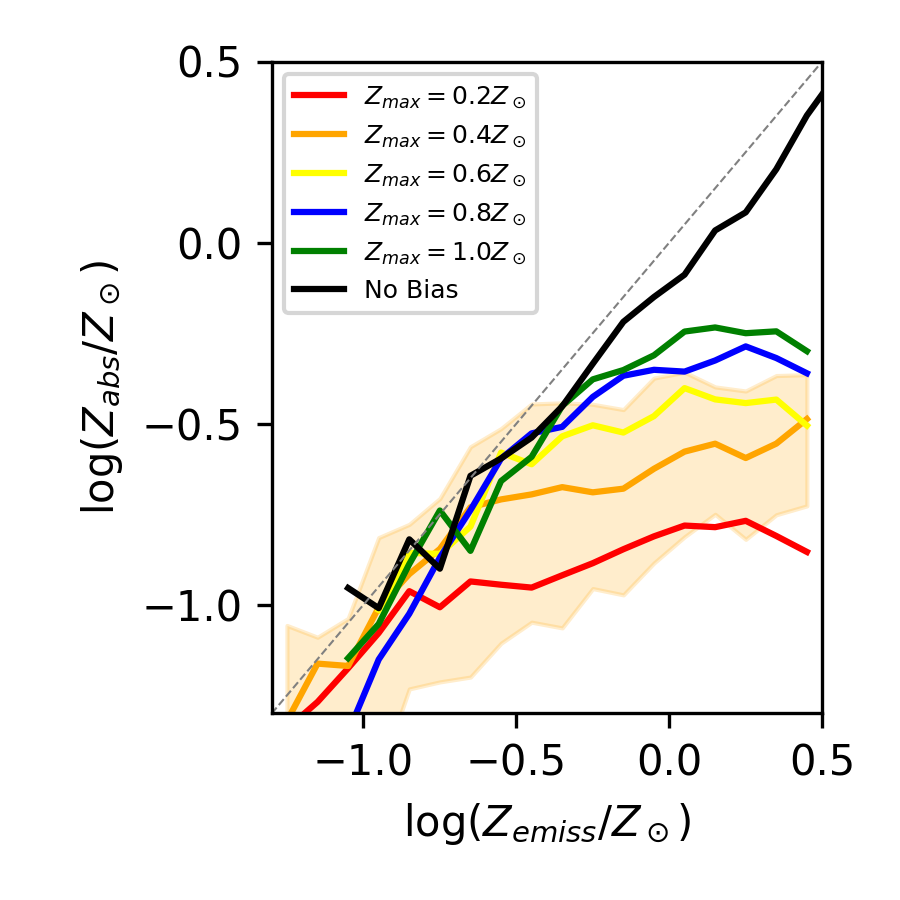}
	\includegraphics[width=0.49\textwidth]{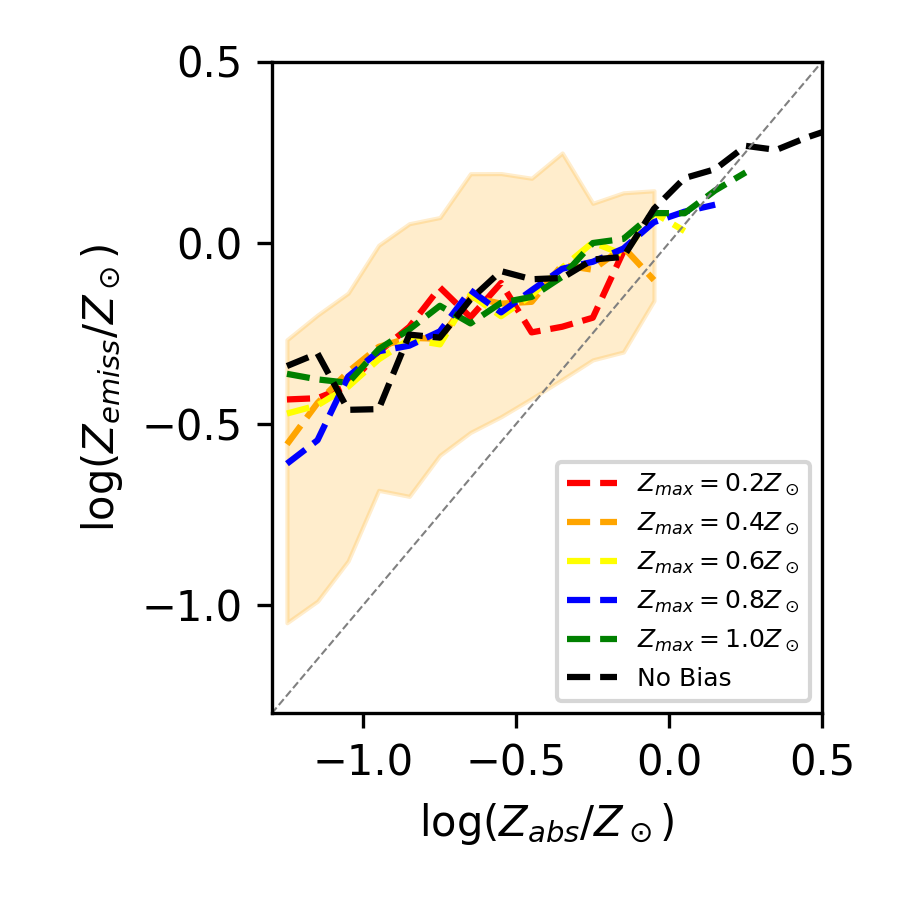}
    \caption{\emph{Left:} The median value of $Z_{\rm abs}$ for TNG galaxies with a fixed value of $Z_{\rm emiss}$. \emph{Right:} The median value of $Z_{\rm emiss}$ for TNG galaxies with a fixed value of $Z_{\rm abs}$. Both of these plots were produced using snap 30 of the TNG100-1 simulation, at a redshift of $z=2.32$. An error region showing the $16$th- and $84$th-percentiles for a GRB metallicity bias function with $Z_{\rm max}=0.4Z_\odot$ is shown in both panels - error regions are of a similar size for all models.}
    \label{fig:z=2.33_plots}
\end{figure*}

For each galaxy, the rate of long GRB formation was computed via equation \ref{eq:cell_wise_rate}, using $100$ different cutoff models with values of $Z_{\rm max}$ between $0.01Z_\odot$ and $1.0Z_\odot$. For each model, at each redshift, 2,000 galaxies were selected (with replacement) to be GRB host galaxies, with the probability of being selected proportional to the rate of GRB formation. For each selected galaxy, one gas cell was chosen to be the site of GRB formation, with the likelihood of each gas cell being chosen proportional to the rate of GRB formation of that gas cell. A random direction for the line-of-sight between that gas cell and the observer was selected, and this was used to compute $Z_{\rm abs}$ for each galaxy.

The theoretical relationship between $Z_{\rm abs}$ and $Z_{\rm emiss}$ for a range of cutoff models at a redshift of $z=2.32$ is shown in Figure \ref{fig:z=2.33_plots}.

In the left panel of Figure \ref{fig:z=2.33_plots}, we show the median value of $Z_{\rm abs}$ for galaxies binned by $Z_{\rm emiss}$, for a selection of GRB metallicity bias models. In this graph, we see that when GRBs are assumed to be unbiased tracers of star formation (solid black line), the expected values of $Z_{\rm abs}$ and $Z_{\rm emiss}$ tend to agree to within $0.1$ dex. The systematic offset between the median values of $Z_{\rm abs}$ and $Z_{\rm emiss}$ can be explained through geometry; in any galaxy, the median radius from the galactic centre where star formation is occurring must be greater than zero. Starting from any point on this radius, there are more lines of sight that point away from the high-$Z$, high-SFR galactic centre than there are that pass through it; therefore, even when no metallicity cutoff is imposed, we see that $Z_{\rm abs}$ is expected to be slightly lower than $Z_{\rm emiss}$.

When GRBs are restricted to form in cells with metallicity below a given threshold, we see that $Z_{\rm abs}$ is predicted to be much lower than $Z_{\rm emiss}$, especially for higher metallicity galaxies. This is essentially by construction. When $Z_{\rm emiss}$ is significantly below  the threshold metallicity for GRB formation, almost all star-forming gas cells in the galaxy have metallicities below $Z_{\rm max}$, and so GRBs act as unbiased tracers of star formation in these galaxies. In these situations, the expected value of $Z_{\rm abs}$ tends to agree with $Z_{\rm emiss}$, mirroring the case where there is no metallicity bias for GRB formation.
However, when $Z_{\rm emiss}$ becomes comparable to $Z_{\rm max}$, we see $Z_{\rm abs}$ become significantly lower than $Z_{\rm emiss}$ for the same galaxies. This turn-off can be explained by noticing that the highest metallicity gas cells are found in the central regions of TNG galaxies, where the SFR is highest. These central regions will dominate contributions to emission-line flux; however, they will often be avoided by lines-of-sight originating from lower-metallicity regions where GRBs are formed (preferentially located in the outskirts of these systems). The characteristic turn-off shape in the  $Z_{\rm abs}-Z_{\rm emiss}$ relation is seen in all metallicity bias models. If the location of this turn-off could be constrained using observational data, this could shed light on the metallicity bias function for GRB progenitors, which could in turn inform models of GRB formation.

In the right hand panel of Figure \ref{fig:z=2.33_plots}, we show the median value of $Z_{\rm emiss}$ for GRB hosts with a given value of $Z_{\rm abs}$. In this situation, we see the same relationship between $Z_{\rm emiss}$ and $Z_{\rm abs}$ for all GRB metallicity bias functions. This indicates that this relation is not dependant on the GRB metallicity bias function. Rather, this statistic depends on the chemical inhomogeneities within TNG galaxies. The difference between $Z_{\rm abs}$ and $Z_{\rm emiss}$ seen in this plot reflects the fact that low-metallicity GRB progenitors are often found in the low-metallicity outer regions of TNG galaxies, where randomly oriented lines of sight are more likely to trace through low-metallicity regions; for galaxies without metallicity gradients, $Z_{\rm abs}$ and $Z_{\rm emiss}$ would be expected to agree.

For all GRB metallicity bias models, the relationship giving the median value of $Z_{\rm emiss}$ for any given value of $Z_{\rm abs}$ is well approximated by a simple power law:

\begin{equation}
    \log (Z_{\rm emiss} / Z_\odot) \approx 0.4 \cdot \log (Z_{\rm abs} / Z_\odot)
    \label{eq:lin_reg_em}
\end{equation}

This equation can be viewed as a prediction of the IllustrisTNG simulation. 
By collecting a large enough sample of GRB host galaxies for which both $Z_{\rm abs}$ and $Z_{\rm emiss}$ are known, predictions of this equation can be compared to observational data as a novel test of the subgrid chemical enrichment schemes used in IllustrisTNG. Similar predictions could also be developed using other cosmological simulations.

In both panels of Figure \ref{fig:z=2.33_plots}, the shaded $1 \sigma$ confidence interval for the GRB formation model with $Z_{\rm max} = 0.4Z_\odot$ are also shown. Error regions are of similar size for other models. The median size of this error region is $^{+0.18}_{-0.27}$ dex for the left hand panel, and $^{+0.34}_{-0.30}$ dex for the right hand panel. This large scatter comes about because measured values of $Z_{\rm abs}$ generally vary substantially for the same galaxy, depending on the position of the GRB within the galaxy, and the orientation of the line-of-sight between the site of the burst and the observer. For each value of $Z_{\rm max}$, we compute the Pearson product moment correlation between $Z_{\rm emiss}$ and $Z_{\rm abs}$, finding $\rho = 0.25-0.4$ across the various cases, and a trend of $\rho$ decreasing
as $Z_{\rm max}$ increases, which mirrors increased departures from linearity for lower GRB-formation threshold metallicities.

At first glance, the differences between these two statistics are counter-intuitive. It is not trivial that the median value of $Z_{\rm emiss}$ for galaxies with a given value of $Z_{\rm abs}$ should differ to the median value of $Z_{\rm abs}$ for galaxies with a given value of $Z_{\rm emiss}$. To understand and explain this result, we turn to the full two-dimensional distributions between $Z_{\rm abs}$ and $Z_{\rm emiss}$.

\begin{figure*}
    \includegraphics[width=0.98\textwidth]{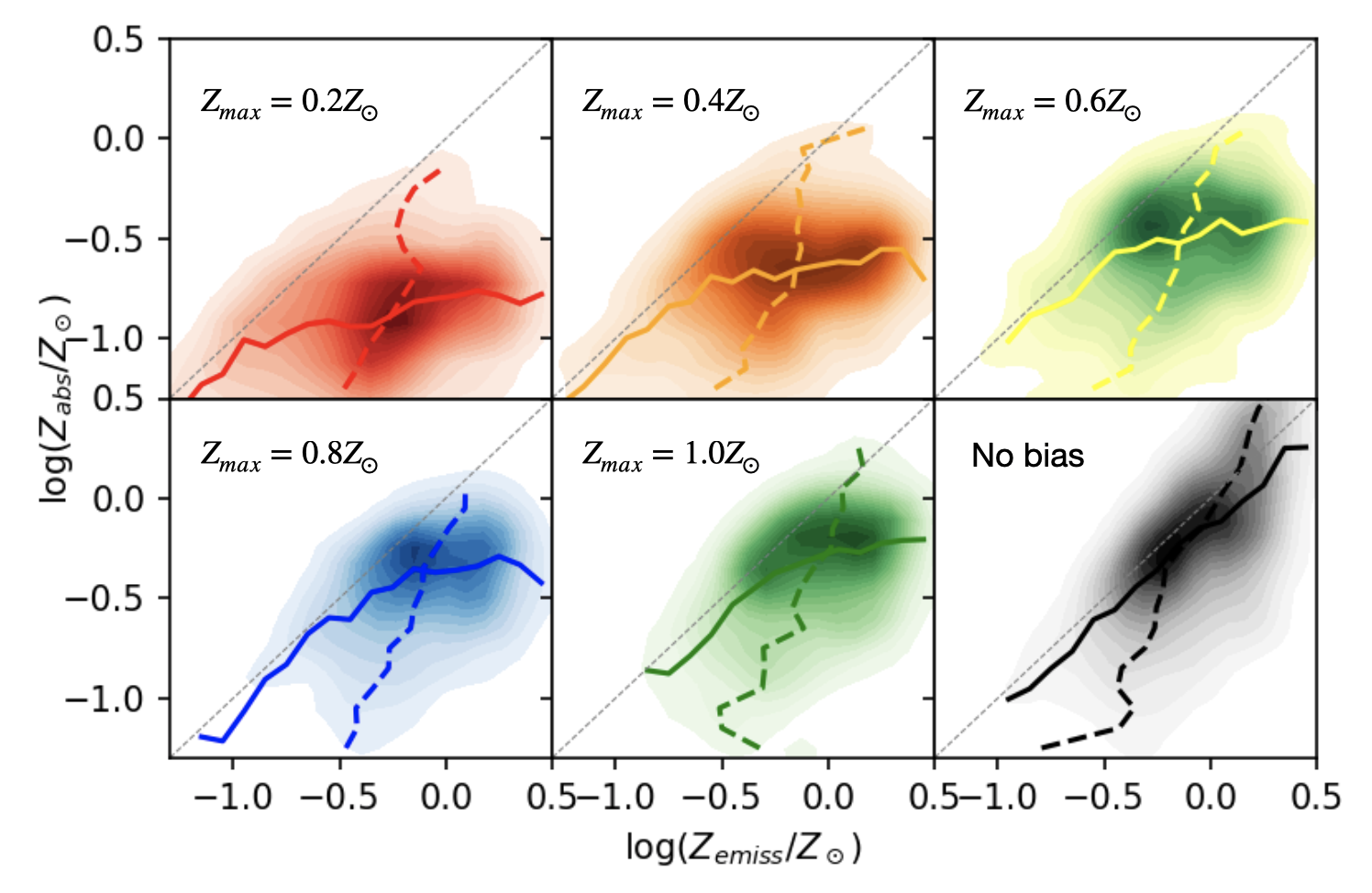}
    \caption{Kernel density estimate plots showing the relationship between $Z_{\rm abs}$ and $Z_{\rm emiss}$ for GRB host galaxies in the IllustrisTNG simulation at a redshift of $z=2.33$, using a variety of GRB metallicity bias functions. From top left to bottom right: $Z_{\rm max} = 0.2Z_\odot$, $Z_{\rm max} = 0.4Z_\odot$, $Z_{\rm max} = 0.6Z_\odot$, $Z_{\rm max} = 0.8Z_\odot$, $Z_{\rm max} = 1.0Z_\odot$, and assuming GRBs are unbiased tracers of star formation. The median values of $Z_{\rm abs}$ for fixed values of $Z_{\rm emiss}$ are shown as solid lines in each plot. Similarly, median values of $Z_{\rm emiss}$ for fixed values of $Z_{\rm abs}$ are shown as dotted lines.} 
    \label{fig:Zabs-Zem-distributions}
\end{figure*}
Figure \ref{fig:Zabs-Zem-distributions} shows these distributions for the metallicity bias functions considered in Figure \ref{fig:z=2.33_plots}. On top of each distribution, the median values of $Z_{\rm emiss}$ for galaxies binned by $Z_{\rm abs}$ are plotted as dashed lines, and the median values of $Z_{\rm abs}$ for galaxies binned by $Z_{\rm emiss}$ are plotted as solid lines\footnote{These solid and dotted lines are computed in the same way as the lines in Figure \ref{fig:z=2.33_plots} -- any differences simply arise from different Monte Carlo sampling.}. We see that, for every metallicity cutoff, there are very few GRB hosts with metallicities lower than $\log(Z_{\rm emiss} / Z_\odot) = -0.6$. Galaxies with very different values of $Z_{\rm emiss}$ may still have the same values of $Z_{\rm abs}$. For this reason, the median value of $Z_{\rm emiss}$ for each bin of galaxies with similar values of $Z_{\rm abs}$ will be drawn closer to the median value of $Z_{\rm emiss}$ of all galaxies. 

\begin{figure*}
    \includegraphics[width=0.7\textwidth]{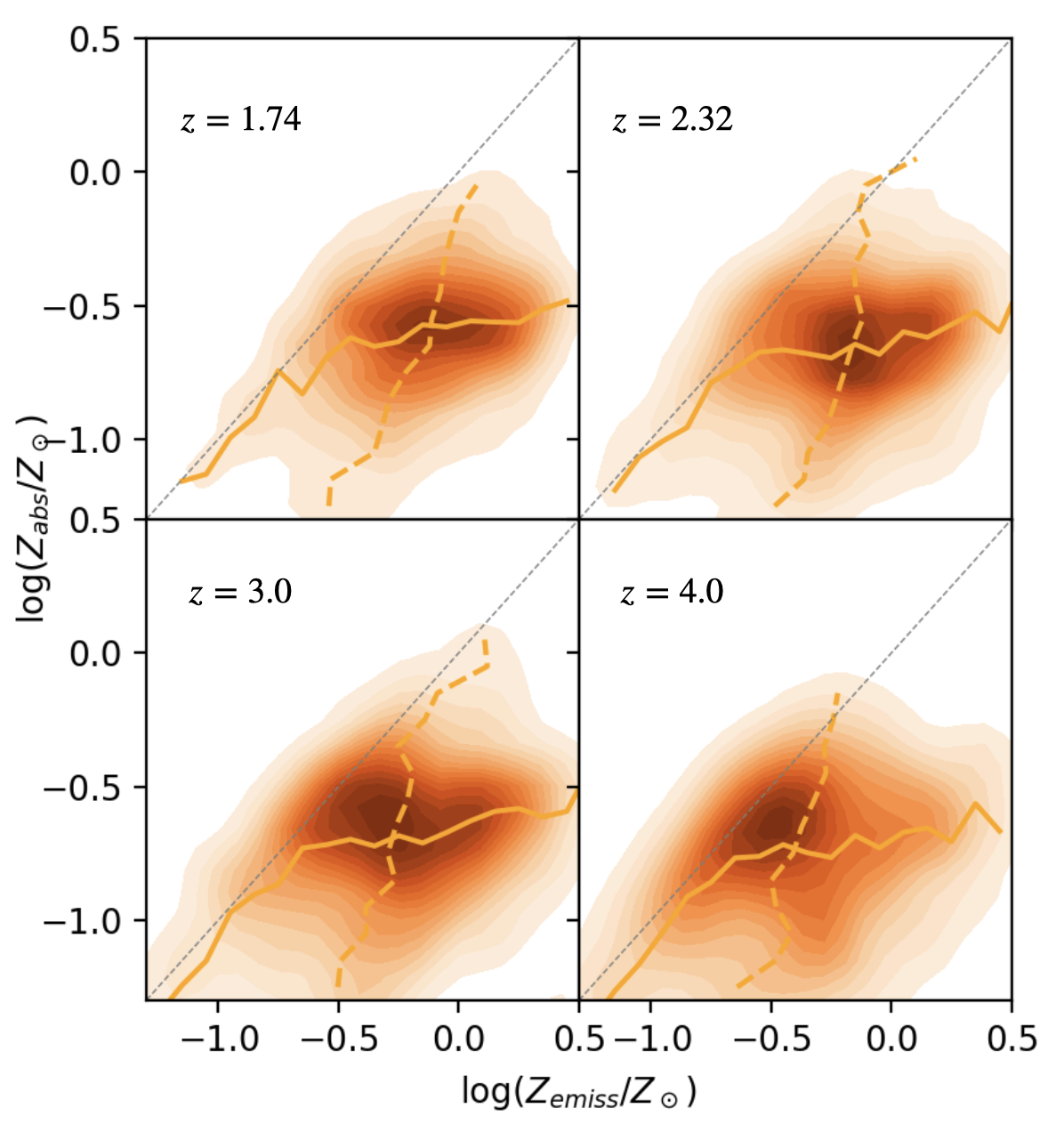}
    \caption{The $Z_{\rm abs}-Z_{\rm emiss}$ relation at a variety of redshifts, assuming a GRB metallicity bias function of $Z_{\rm max} = 0.4 Z_\odot$. While the median value of $Z_{\rm emiss}$ for GRB hosts decreases as redshifts increase, the median value of $Z_{\rm abs}$ for galaxies with a fixed $Z_{\rm emiss}$ shows no redshift dependence. Trends are similar for other GRB metallicity bias models.}
    \label{fig:redshift-independance}
\end{figure*}

As an additional feature, we note that for any fixed value of $Z_{\rm max}$, the $Z_{\rm abs}-Z_{\rm emiss}$ relation showed no statistically significant variation with redshift for $1.8\leq z \leq 4$. In Figure \ref{fig:redshift-independance}, we plot the full distribution of $Z_{\rm abs}$ and $Z_{\rm emiss}$ for a GRB formation model with $Z_{\rm max} = 0.4 Z_\odot$ at a range of redshifts; similar results were seen for all values of $Z_{\rm max}$ that were explored. As $z$ is increased, the median $Z_{\rm emiss}$ of GRB host galaxies decreases, slightly decreasing the median values of $Z_{\rm emiss}$ for galaxies of a fixed $Z_{\rm abs}$. However, the median value of $Z_{\rm abs}$ for GRB hosts with a fixed value of $Z_{\rm emiss}$ does not change with redshift. 
At higher $z$, the spread in $Z_{\rm abs}$ increases to include more low metallicity values. This is because at lower redshifts ($z\sim 2$), gas cells have higher metallicities overall, and so there is a smaller range in the metallicities of gas cells with $Z_{\rm cell} < Z_{\rm max}$.

\section{Observational Opportunities} \label{sec:observations}

\subsection{Current data}
\label{subsec:GRB121024A}

At time of writing, GRB121024A is the only GRB whose host galaxy has been determined both using absorption-line and emission-line methods \citep{Arabsalmani+18}. 
%Using the DLA method, \citet{Cucchiara+15} determined the metallicity of the host galaxy of this event to be $\log(Z/Z_\odot) = -0.4 \pm 0.12$.

The emission-line metallicity was first measured by \citet{Friis+15} and then revised by \citet{Kruhler+15}, who quote a value of $12+\log ([O/H]) =8.41^{+0.11}_{-0.12}$. This value was obtained using strong-line metallicity diagnostics calibrated from local Universe observations. One challenge with measuring gas metallicities at high-redshift with emission lines is that conditions in the emitting regions likely evolve with redshift \citep{Kewley+13, Steidel+14, Strom+17}. Recently, a number of studies have suggested that high-redshift metallicity measurements may be systematically offset from standard local Universe calibrations \citep[e.g.][]{Bian+18, Sanders+20}. On this basis, we re-derive the emission line metallicity using the emission line data of \citet{Kruhler+15} and the NebulaBayes fitting package \citep{Thomas+18}, obtaining a value of $12+\log ([O/H])=8.36^{+0.12}_{-0.24}$, slightly lower than but consistent with \citet{Kruhler+15}. The reader is referred to Appendix \ref{ap:gas_metallicity} for a more detailed discussion of the challenges associated with measuring metallicity from emission lines at high-redshift, and the justification behind our revised value quoted above.
Taking the value of solar metallicity to be $12+\log([O/H])_\odot = 8.69$ \citep{Asplund09}, our $Z_{\rm emiss}$ value corresponds to a value of $\log(Z/Z_\odot) = -0.33^{+0.12}_{-0.24}$.

Comparing the emission line metallicity of the host galaxy of GRB121024A with observational samples from high-redshift surveys is highly dependent on the metallicity diagnostic used as well as how it is calibrated (refer to discussion in Appendix A).
Using a local Universe calibration of the O3N2 diagnostic \citep{PettiniPagel04}, we find that the host galaxy of GRB121024A is highly consistent with $z\sim2.3$ mass-metallicity relation measurements from both KBSS \citep{Steidel+14} and MOSDEF \citep{Sanders15} samples.
More recently, \citep{Sanders+20b} re-evaluated the $z\sim2.3$ mass-metallicity relation in MOSDEF employing a redshift-dependent metallicity calibration.
Based on the high-redshift analogue sample of \citet{Bian+18}, this calibration suggest a somewhat higher value for galaxies of the mass of the host of GRB121024A (12+log(O/H)$\sim$8.48, rather than 12+log(O/H)$\sim$8.20).
While the \citet{Kruhler+15} observations do not measure all of the same emission lines used in that study, our measured metallicities using \citet{Bian+18} calibrations of R23, O32, and O3N2 diagnostics for this galaxy are all around $\sim$0.1 dex lower than the expected value from the \citep{Sanders+20b} $z\sim2.3$ mass-metallicity relation.
GRB hosts at this redshift are expected to more or less trace the star-forming population, with a slight bias to lower mass, lower metallicity galaxies \citep{Palmerio+2019}. Thus, the comparison we find between the host of GRB121024A and $z\sim2.3$ observational samples appears to match with expectations.

The absorption-line metallicity for the host of GRB121024A was first determined by \citet{Cucchiara+15} to be $\log(Z_{\rm abs}/Z_\odot) = -0.4 \pm 0.12$. Recently, the result was revisited by \citet{Bolmer+19} with an improved modeling of dust. Using spectral data for 22 GRB afterglows detected with \emph{VLT}/X-SHOOTER, \citet{Bolmer+19} determined the relative strength of a collection of metal absorption lines using Markov-Chain Monte Carlo methods, simultaneously fitting for the broadening parameter, and the continuum flux of the GRB. Values of $\log (Z_{\rm abs} / Z_\odot ) $ were computed using the normalisations of \citet{Asplund09}. The effects of intervening dust were accounted for by using calibration curves based on \citet{DeCia16}. Given these recent data analysis improvements, we  take the value of $Z_{\rm abs}$ for the host of GRB121024A to be $\log (Z_{\rm abs}/Z_\odot) = -0.68 \pm 0.07$, as determined by \citet{Bolmer+19}. Notably, this result is significantly lower than the value of $Z_{\rm emiss}$ for the galaxy. Rather than suggesting possible observational and/or data analysis/calibration issues related to these metallicity determinations, we highlight that the discrepancy can be naturally accounted if we simply consider the scenario where the lower metallicity seen in the spectra of GRB121024A is a result of the GRB forming in a low metallicity environment, as predicted by the models we consider here.

\begin{figure*}
	\includegraphics[width=0.49\textwidth]{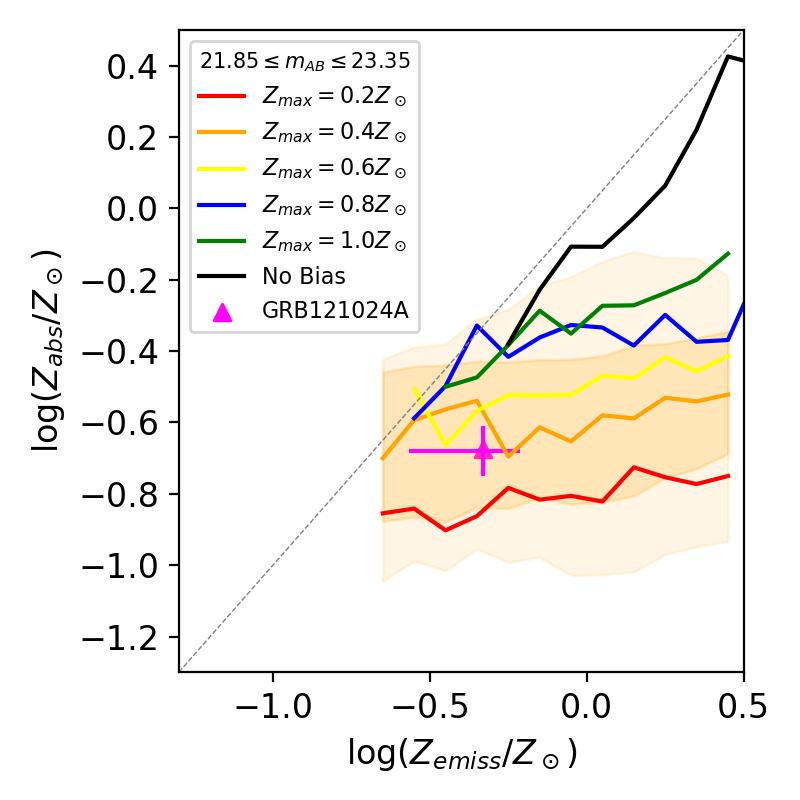}
	\includegraphics[width=0.49\textwidth]{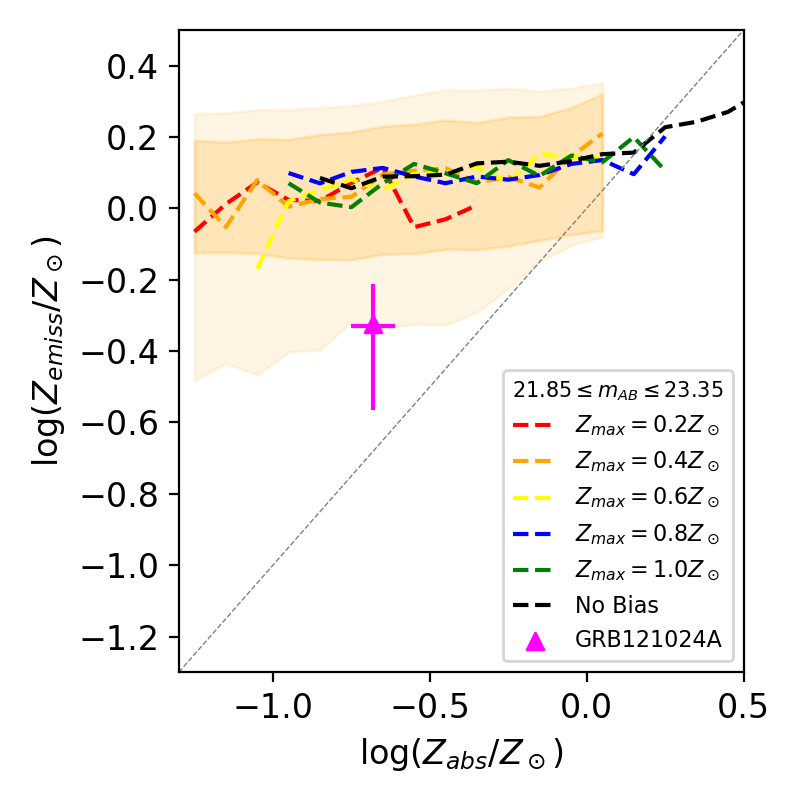}
    \caption{Same as Figure \ref{fig:z=2.33_plots}, but plotting only TNG galaxies with brightness within $0.75$ magnitudes of the host galaxy of GRB121024A. Values of $Z_{\rm abs}$ and $Z_{\rm emiss}$ for the host galaxy of GRB121024A are shown in magenta. Orange shaded regions represent the 68\% (dark) and 90\% (light) confidence intervals for the model with $Z_{\rm max}=0.4Z_\odot$.}
    \label{fig:mag_limited}
\end{figure*}

In Figure \ref{fig:mag_limited}, we plot the $Z_{\rm abs}$ and $Z_{\rm emiss}$ of the host of GRB121024A superimposed on the curves that show the theoretical $Z_{\rm abs}-Z_{\rm emiss}$ relations for a variety of simple GRB metallicity bias cutoff functions. Due to the substantial brightness of this host ($L>L_*$), we only compare it to similar galaxies in the IllustrisTNG simulation, defined as those with rest-frame U-band magnitudes between $21.85$ and $23.35$. The magnitude of each host galaxy in the TNG simulation is determined from the total brightness of all sub-halo components that we associate to a single galaxy (see \citealt{Paper1}). The $U$-filter bandpass is from \citet{Buser78}. Flat SEDs in frequency and a flat cosmology with parameters chosen to match \cite{Planck16} are assumed for the purpose of computing K-corrections and distance modulus.

Looking at the left panel of Figure \ref{fig:mag_limited}, this data point seems to agree with the IllustrisTNG models for which $ Z_{\rm max} \sim 0.4 Z_\odot$, and disagrees with the GRB formation model with no metallicity bias. However, in the right panel, we see that this data point has a lower value of $Z_{\rm emiss}$ than is expected from the TNG simulation for any GRB metallicity bias model. Shaded regions over both plots indicate 68\%  and 90\% confidence intervals for the model in which $Z_{\rm max} = 0.4Z_\odot$, calculated using a running mean across $Z_{\rm emiss}$-- error regions for other models are of a comparable size. The host galaxy of GRB121024A lies outside of this 68\% confidence interval, but it does lie at the edge of the 90\% confidence interval. Thus, we conclude that the limited data-model comparison highlights a qualitative off-set between absorption and emission metallicities, and the host galaxy of GRB121024A has an emission-line metallicity that may be in slight tension with the $Z_{\rm emiss}$ distribution of comparably bright TNG galaxies at the same redshift.

In order to determine which GRB metallicity bias model is most likely to be responsible for the formation of GRB121024A, it is not sufficient to simply compare this data point to the median $Z_{\rm abs}-Z_{\rm emiss}$ curves in Figure 4.
Instead, we determine the relative likelihood of each cutoff model, using the following general equation based on the law of total probability. 

Let $\vec{x_o}$ be a vector of observable parameters. Let $p_{sim}(\vec{x})$ be the multivariate probability density function of $\vec{x}$ determined from a given theory/numerical model. Then, the probability of measuring $\vec{x_o}$ assuming that the model is correct is given by:

\begin{equation}
    \label{eq:general_likelihood}
    Pr(\text{measure } \vec{x_o} \, | \text{ sim}) = \int Pr(\text{measure } \vec{x_o} \,|\, \vec{x}) p_{sim}(\vec{x}) d\vec{x}
\end{equation}

Because values of $Z_{\rm abs}$ and $Z_{\rm emiss}$ for this galaxy were measured using different instruments and different pipelines, the two measurements are independent, and therefore any errors associated with these measurements are uncorrelated.
For this reason, we can separate the bivariate distribution $Pr(\text{measure } \vec{x_o} \,|\, \vec{x})$ into the product of two univariate distributions -- one for $\log(Z_{\rm emiss}/Z_\odot)$, and one for $\log (Z_{\rm abs}/Z_\odot)$. For simplicity, we assume that $\log (Z_{\rm abs}/Z_\odot)$ follows a normal distribution, with a mean of $-0.4$ and a standard deviation of $0.11$. The distribution of $\log (Z_{\rm emiss}/Z_\odot)$ is taken from the marginalised probability density function of metallicity output by NebulaBayes (see the top left panel of Figure \ref{fig:ap_NB_post}).

Using this method, the relative likelihood for $100$ cutoff models with values of $Z_{\rm max}$ ranging from $0.01 Z_\odot$ to $1.0 Z_\odot$ was computed. The most likely model was found to have $Z_{\rm max} = 0.31 Z_\odot$; however, the likelihood function was fairly flat, with all models with $0.26Z_\odot \leq Z_{\rm max} \leq 0.41 Z_\odot$ having relative likelihoods greater than $0.9$, and all models with $0.16Z_\odot \leq Z_{\rm max} \leq 0.87 Z_\odot$ having relative likelihoods greater than $0.5$. %and all models with $0.24 \leq Z_{\rm max}/Z_\odot \leq 0.52$ having relative likelihoods greater than $0.8$.
Metallicity bias functions with cutoff metallicities of $0.07 Z_\odot$ or less had relative likelihoods below $0.135$, allowing them to be excluded with 95\% confidence \citep{EasyStats}. From this, we conclude that this single data point is not sufficient to constrain the strength of the GRB metallicity bias function. Instead, a larger sample of GRB hosts for which both $Z_{\rm abs}$ and $Z_{\rm emiss}$ are known is required.

\subsection{Future data opportunities}
\label{subsec:jwst_opportunities}

GRB-DLA observations are only possible at $z \gtrsim 1.8$ with current telescopes; this redshift is the minimum necessary so that the Lyman-$\alpha$ trough is redshifted to a wavelength where it is visible through the atmosphere \citep{Cucchiara+15}. Because of this, GRB hosts with metallicities measured via the GRB-DLA method are often too faint to be followed up on with ground-based spectroscopy.

\begin{table*}
    \centering
    \begin{tabular}{|c|c|c|c|c|c|c|}
         \hline
          GRB ID& Redshift &$\log (Z_{\rm abs}/Z_\odot )$ &$m_{AB}$ & Wavelength $ (\mu$m) & Reference  & $A_V$ (mag)\\
         \hline
000926A & 2.04 & $0.17 \pm 0.34^{b}$ & $24.59 \pm 0.01$ & $0.289$ & (1)& $0.38 \pm 0.051$ (i) \\
050401A & 2.90 & $-0.92 \pm 0.68^{b}$ & $24.29 \pm 0.39$ & $3.6$ & (2) & $0.45 \pm 0.035$ (ii)\\
050730A & 3.97& $-2.31 \pm 0.18^{b}$ & $>25.47$ & $3.6$ & (2) & — \\
050820A & 2.62   &$-0.49 \pm 0.10^{b}$ & $25.3 \pm 0.3$ & $1.63$ & (3) & $0.14$ (iii)\\ 
050922C & 2.20   &$-1.88 \pm 0.14^{a}$& $>25.34$ & $3.6$ & (2)& — \\ 
070802A & 2.45 & $-0.24 \pm 0.80^{b}$ & $21.70 \pm 0.25$ & $2.2$ & (4)& $1.23 \pm 0.05$ (iv) \\ 
080804A & 2.21   &$-0.75 \pm 0.13^{a}$& $24.76 \pm 0.23 $ & $3.6$ & (2) & $0.05$ (iii)\\
081008A & 1.96   &$-0.51 \pm 0.17^{b}$& $>24.75$ & $3.6$ & (2)& — \\ 
090323A & 3.57 & $-0.41\pm 0.11^{b}$ & $24.25 \pm 0.18$ & $0.764$ & (5) & $0.10 \pm 0.04$ (ii)\\
090926A & 2.11 &$-1.72 \pm 0.05^{c}$ & $>24.4$ & $0.476$ & (6) & —  \\
100219A & 4.67 &$-1.16 \pm 0.11^{c}$ & $26.6 ^{+ 0.8}_{-0.4} $ & $0.78$ & (7) & $0.15^{+0.04}_{-0.05}$ (v)\\
111008A & 4.99 &$-1.79 \pm 0.10^{c}$ & $>25.6$ & $0.91$ & (8) & —  \\
120119A & 1.73 & $ -0.79 \pm  0.42^{b}$ & $22.89 \pm 0.07$ & $3.6$ & (2) & $1.06 \pm 0.02$ (vi) \\
120327A& 2.81&$-1.34 \pm 0.12^{c}$ & $24.50 \pm 0.23$ & $0.63$ & (9) & $0.05 \pm 0.02$ (vii) \\ 
120909A & 3.93 &$-0.29 \pm 0.10^{c}$ & $24.95 \pm 0.12$ & $0.768$ & (10) & $ 0.16 \pm 0.04$ (vii)\\ 
121024A & 2.30 &$-0.68 \pm 0.07^{c}$ & $22.6 \pm 0.2 $ & $2.146$ & (11) & $0.26 \pm 0.07$ (vii) \\
130408A & 3.76 &$-1.46 \pm 0.05^{c}$ & $>25.9$ & $0.622$ & (10)& — \\ 
         \hline
090809A & 2.74 &$-0.46 \pm 0.15^{c}$ & — & —  & & —  \\ 
100425A & 1.76   &$-0.96 \pm 0.42^{a}$ & — & — &  & — \\
111107A & 2.89 &$-0.28 \pm 0.15^{c}$ & — & — & &  — \\
120716A & 2.48 &$-0.57 \pm 0.08^{c}$ & — & — & & —  \\
120815A & 2.36 &$-1.23 \pm 0.03^{c}$ & — & — & & — \\ 
130606A & 5.91 &$-1.58 \pm 0.08^{c}$ & — & — & & — \\ 
140311A & 4.96 &$-2.00 \pm 0.11^{c}$ & — & — & & — \\ 
141028A & 2.33 &$-1.62 \pm 0.28^{c}$ & — & — & & — \\
141109A & 2.99 &$-1.37 \pm 0.05^{c}$ & — & — & & — \\ 
150403A & 2.06 &$-0.02 \pm 0.05^{c}$ & — & — & & — \\ 
151021A & 2.33 &$-0.97 \pm 0.07^{c}$ & — & — & & — \\ 
151027B & 4.07 &$-0.59 \pm 0.27^{c}$ & — & — & & — \\ 
160203A & 3.52 &$-0.92 \pm 0.04^{c}$ & — & — & & — \\
161023A & 2.71 &$-1.05 \pm 0.04^{c}$ & — & — & & — \\
170202A & 3.65 &$-1.02 \pm 0.13^{c}$ & — & — & & — \\ 
    \end{tabular}
    \caption{Photometric properties of the 32 GRB host galaxies for which $Z_{\rm abs}$ has been measured. References for magnitudes: (1) \citet{Castro+03}; (2) \citet{SHOALS2}; (3) \citet{Chen+09}; (4) \citet{Eliasdottir+09}; (5) \citet{McBreen+10}; (6) \citet{Cenko+11}; (7) \citet{Thone+13};  (8) \citet{Sparre+14}; (9) \citet{D'Elia+14}; (10) \citet{Griener+15}; (11) \citet{Friis+15}. 
    References for metallicities: $(a)$ \citet{Cucchiara+15}; $(b)$ \citet{Wiseman+17}; $(c)$ \citet{Bolmer+19}. References for dust extinction: (i) \citet{Starling+07}; (ii) \citet{Schady+11}; (iii) \citet{Schady+12}; (iv) \citet{Greiner+11}; (v) \citet{Bolmer+18}; (vi) \citet{Wiseman+17}; (vii) \citet{Bolmer+19}. }
    \label{tab:mangitudes}
\end{table*}

To quantify this statement and assess opportunities for future progress, we searched the literature for magnitudes of the 33 GRB hosts for which $Z_{\rm abs}$ is presented in \citet{Cucchiara+15},  \citet{Wiseman+17}, and/or \citet{Bolmer+19}. Of these galaxies, 17 have measured magnitudes or magnitude limits at IR/optical wavelengths, reported in Table \ref{tab:mangitudes}.
We see that all galaxies except for the hosts of GRB121024A and GRB120119A have magnitudes close to or fainter than 25, making ground-based spectroscopy extremely challenging with current 10m class facilities. 

The \textit{James Webb Space Telescope} (\emph{JWST}) is currently expected to launch in October 2021 \citep{NasaPresser}. For the galaxies in Table \ref{tab:mangitudes} ($z \sim 2-5$), rest-frame optical emission lines typically used for determining metallicity will be redshifted to the near-infrared, making them ideal candidates for the unique NIR capabilities of the \emph{JWST}/NIRSpec fixed slit spectrograph. For each host galaxy in Table \ref{tab:mangitudes} for which the magnitude is known, we use the \emph{JWST} exposure time calculator\footnote{\url{jwst.etc.stsci.edu}} to assess the feasibility of measuring $Z_{\rm emiss}$ using an observation window of a few hours.

Each GRB host is modeled with a SED similar to the starburst galaxy Il ZW 096, as measured by \citet{Brown+14}, rescaled and redshifted as appropriate. We also assume that all galaxies are extended objects, with a Sersic profile with index 1, semi-major axis of 0.5 arcseconds, and semi-minor axis of 0.25 arcseconds. Dust extinction is accounted for using values from Table \ref{tab:mangitudes}, assuming a Milky-way extinction profile with $R_V=3.1$.\footnote{These dust extinction magnitudes were estimated from extinction of the GRB afterglow, and therefore are sensitive only to dust obstruction along the path of the burst, which may not necessarily reflect the dustiness of the host galaxy as a whole. Nevertheless, we adopt these values as the extinction values for the galaxies in question, with this caveat in mind.}
All backgrounds are assumed to be the median background at the celestial positions of the hosts.
We emphasise that each of these calculations are estimates, based on current expectations of \emph{JWST}'s performance abilities informed by data from ground measurements and calibrations.

Of the 11 GRB host galaxies presented in Table \ref{tab:mangitudes} for which the magnitude of the host is known, three are bright enough for $Z_{\rm emiss}$ to be measured by \emph{JWST} with an exposure time of less than eight hours.
The brightest of these is the host of GRB 121024A. We find that after a 30 minute exposure using 2 integrations, using a medium resolution grism and a filter centred at 1.4 $\mu$m, the [O {\sc ii}], [O {\sc iii}] and H$\beta$ spectral lines can all be resolved with a SNR $> 5$, allowing $Z_{\rm emiss}$ to be computed for this galaxy via the $R_{23}$ calibration diagnostic \citep{Kewley+Ellison08}. As this galaxy has already has $Z_{\rm emiss}$ measured using the \emph{VLT}/X-SHOOTER NIR instrument, the utility of a second emission metallicity measurement for this galaxy is limited. For this reason, we investigated the possibility of using NIRSpec's integral field unit in order to measure the metallicity distribution of this host. We found that using an exposure time of 1.5 hours and a Nod-In-Scene strategy to remove backgrounds, metallicity measurements for five different regions of this galaxy could be resolved independently, allowing for crude estimates of the metallicity gradient of this host.

With an exposure time of $\sim$1 hour, several strong emission lines of the host galaxy of GRB120327A are predicted to be measurable with S/N $\gtrsim 4$, including the lines for [O {\sc ii}], [N {\sc ii}], H$\alpha$ and H$\beta$. This would allow $Z_{\rm emiss}$ to be computed for this host via the O$3$N$2$ diagnostic. Similarly, using an exposure time of $\sim$1 hour and a medium resolution grism between 1.6$\mu$m and 3.0$\mu$m would allow detections of the H$\alpha$, H$\beta$, [O {\sc ii}], [O {\sc iii}], and [N {\sc ii}] lines for the host galaxy of GRB090323A with S/N $\gtrsim 4$, allowing $Z_{\rm emiss}$ to be detemined with any of the $R_{23}$, O$3$N$2$, or N$2$O$2$ diagnostics \citep{Kewley+19}.

With an exposure time of 2h, four key emission lines of the host galaxy of GRB000926A can be seen with S/N $> 3$, allowing $Z_{\rm emiss}$ to be measured via the $R_{23}$ diagnostic. Similarly,  [O {\sc iii}], H$\alpha$, [N {\sc ii}], and  H$\beta$ lines would be visible for the host galaxy of GRB070802A with a S/N $> 4$ using a $\sim$2 hour observation window, allowing $Z_{\rm emiss}$ to be determined using the O$3$N$2$ diagnostic.

In addition to these five systems, we report two more that are liminally visible, requiring \emph{JWST} observations of $\sim$4.5 hours in order for all key lines to be observed with S/N$\gtrsim 3$.\footnote{Single exposures using \emph{JWST} are limited to $10000$ seconds, or $\sim 2.8$ hours. These targets would therefore require multiple exposures in order for $Z_{\rm emiss}$ measurements to be made.}
Using a 4.5 hour observation window, a medium resolution grism and a filter centered at $2.4 \mu$m, the high-redshift host galaxy of GRB120909A could have its metallicity obtained via the $R_{23}$ diagnostic, with S/N $\gtrsim 3$ for all necessary emission lines (S/N = 3.03 for [O {\sc ii}], 5.22 for [O {\sc iii}], 3.26 for H$\beta$).
While the host galaxy of GRB120119A at $z=1.73$ is very bright ($m_{AB} = 22.89\pm0.07$; comparable to the host galaxy of GRB121014A for which $Z_{\rm emiss}$ is known), it is also highly dust-obscured, with $A_V=1.06\pm0.02$. This dust increases the difficulty of measuring strong line ratios for this galaxy -- however, with an exposure time of 4.5 hours, this metallicity can likely be detected using the O$3$N$2$ diagnostic (S/N = $2.78$ for H$\beta$, $4.13$ for [O {\sc iii}], $7.95$ for H$\alpha$, $3.03$ for [N {\sc ii}]).

We found that all other hosts with known magnitudes are too faint for a sufficient collection of emission lines to be detected with SNR $\gtrsim 2$ for total exposure times of less than $8$ hours. Of the remaining hosts, six were too faint to be detected in the existing (deep) imaging observations, so it is unlikely that they will be bright enough for $Z_{\rm emiss}$ measurements via NIRSpec. For the other 15 GRB hosts with known absorption metallicities, no attempts to detect the host galaxies have been published. Based on sample size, it is likely that some of these galaxies would be bright enough for $Z_{\rm emiss}$ measurements via NIRSpec. In addition to a \emph{JWST} observing campaign, ground-based follow-up attempts to detect these as-yet undetected GRB hosts, for example with \emph{VLT}, could be performed in an effort to increase the number of data points available to constrain the $Z_{\rm abs}-Z_{\rm emiss}$ relation.

% AJC: I added a subsection break. Feel free to revert if you prefer
\subsection{Predictions for $Z_{\rm emiss}$}

By searching for analogous galaxies in the IllustrisTNG simulation, estimates of $Z_{\rm emiss}$ were computed for the $11$ GRB host galaxies for which both $Z_{\rm abs}$ and $m_{AB}$ are known. IllustrisTNG galaxies were selected to (i) be at a similar redshift to the observed galaxies, (ii) have similar magnitudes, (iii) be likely GRB host candidates, and (iv) have similar values of $Z_{\rm abs}$. For each observed host, an ensemble of analogous galaxies were selected from the snapshot with the redshift closest to the measured redshift of the host galaxy. Because the $Z_{\rm abs}-Z_{\rm emiss}$ relation is not sensitive to redshift, choosing galaxies from snapshots with slightly different redshifts to the observed galaxies was found to not significantly affect the resulting $Z_{\rm emiss}$ estimates. 
After accounting for dust attenuation using the extinction law of \citet{ccm89} and $R_V=3.1$, simulated galaxies with $U$-band magnitudes that differed from observed GRB host magnitudes by more than $0.75$ were excluded. 
The rate of GRB formation and $Z_{\rm abs}$ was determined for each galaxy in the simulation assuming $Z_{\rm max} = 0.33 Z_\odot$.

Because $Z_{\rm abs}$ measurements of the host galaxies are uncertain, the probability of each host having a true value of $\log (Z_{\rm abs} / Z_\odot)$ is modelled as a normal distribution, with mean equal to the observed value of $\log (Z_{\rm abs} / Z_\odot)$, and standard error given by Table \ref{tab:mangitudes}. The probability of each simulated galaxy being an analogue for each host was taken to be proportional to the probability of that simulated galaxy hosting a GRB, multiplied by the probability of the host's measured $Z_{\rm abs}$ value matching the $Z_{\rm abs}$ value of the simulated galaxy (following Equation \ref{eq:general_likelihood}). 

\begin{figure*}
    \centering
    \includegraphics[width=0.98\textwidth]{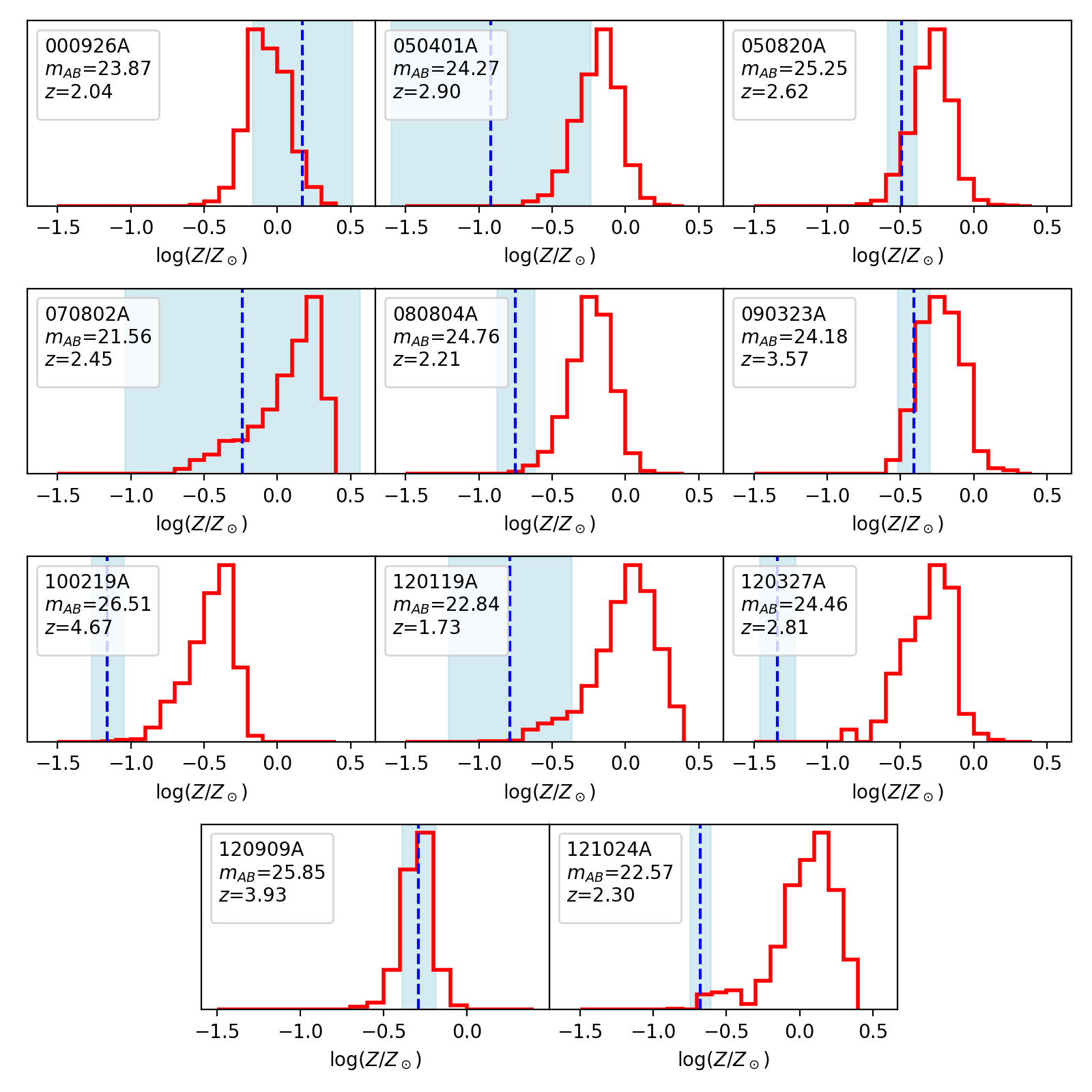}
    \caption{Estimating $Z_{\rm emiss}$ (red lines) for GRB host galaxies with known values of $Z_{\rm abs}$ (dashed blue lines) and known dust-corrected magnitudes, assuming a fiducial cutoff model with $Z_{\rm max} = 0.33 Z_\odot$. For most galaxies, $Z_{\rm emiss}$ is expected to be significantly larger than the measured value of $Z_{\rm abs}$. The shaded blue regions indicate the $1\sigma$ error associated with the $Z_{\rm abs}$ measurements.}
    \label{fig:Z_emis_estimates}
\end{figure*}

Figure \ref{fig:Z_emis_estimates} shows a histogram of the value of $Z_{\rm emiss}$ for each TNG galaxy, weighted by the probability of each TNG galaxy matching the host as described above.
We see that, under this model, the most likely emission metallicity for the host of GRB120327A is approximately $\log (Z_{\rm emiss}/Z_\odot) = -0.25$, substantially higher than the measured $Z_{\rm abs}$ value. This object would thus represent an excellent target for $Z_{\rm emiss}$ observations, since it would have a strong power to falsify/validate our model framework. For other galaxies which are bright enough for \emph{JWST} observation (see Section \ref{subsec:jwst_opportunities}), either the measured value of $Z_{\rm abs}$ for these galaxies is too close to the predicted value of $Z_{\rm emiss}$ (GRB000926A, GRB090323A, GRB120909A) or too poorly constrained (GRB070802A, GRB120119A) to falsify the hypothesis that $Z_{\rm abs}$ and $Z_{\rm emiss}$ are drawn from the same distribution. Finally, we note that for GRB121024A, the expected value of $Z_{\rm emiss}$ under this model is significantly higher than the measured value of $Z_{\rm emiss}$, indicating that this data point is in tension with the IllustrisTNG simulation and/or this fiducial GRB model (see also the data-model comparison in the right panel of Figure \ref{fig:mag_limited}, and associated discussion in the text). 

\section{Discussion} \label{sec:discussion}

Our analysis shows that, in general, $Z_{\rm abs}$ is expected to be lower than $Z_{\rm emiss}$ for GRB host galaxies. It is not, however, correct to say that absorption methods under-predict metallicities of host galaxies; rather, these two methods simply measure different physical quantities. For this reason, we caution the reader about directly comparing metallicities obtained using absorption methods with those obtained via emission-line spectroscopy. A well-established  $Z_{\rm abs}-Z_{\rm emiss}$ theoretical relation is essential in order to compare these properties; and even then, comparisons should only be done over large populations of galaxies when statistical noise can be minimised, as even the same galaxy can have very different values of $Z_{\rm abs}$ (see Figure \ref{fig:stochastic_Z_abs}). 

GRB121024A is currently the only GRB host galaxy for which both $Z_{\rm emiss}$ and $Z_{\rm abs}$ measurements have been made. With the most up to date measurements, we indeed find a systematically higher value for $Z_{\rm emiss}$ than $Z_{\rm abs}$ for this host. In the bottom panel of Figure 4, we showed that, although $Z_{\rm emiss}$ was higher than $Z_{\rm abs}$, it was still below the predicted median values of $Z_{\rm emiss}$ for GRB hosts with our measured $Z_{\rm abs}$ ($\log (Z_{\rm abs} / Z_\odot) $ = $-0.68 \pm 0.07$; we note again however that the $Z_{\rm emiss}$ value is still within 95 \% CI). While it is not possible to draw a strong conclusion from this one data point, it is however worth considering a possible bias toward lower $Z_{\rm emiss}$ galaxies. 

Given the possible presence of observational (sample selection) biases in emission line measurements for GRB host galaxies, it is reasonable to consider we may be biased toward achieving successful measurements in hosts with the highest star-formation rates (and hence the brightest emission lines). 
An anti-correlation between $Z_{\rm emiss}$ and SFR at fixed stellar mass has been well documented in star-forming galaxies \citep[e.g.][]{Ellison+08, LaraLopez+10, Mannucci+10}. This is often referred to as the Fundamental Metallicity Relation (FMR) which has been shown by some authors to be constant out to redshifts of $z\sim3.3$ \citep[e.g.][]{Mannucci+10, Belli+13, Sanders+20}, however other authors have found conflicting results in high-redshift samples \citep[e.g.][]{Yabe+12, Cullen+14, Wuyts+16}. The degree of possible FMR evolution aside, there is a strong possibility that being biased toward detections in galaxies with the highest SFR may bias us toward galaxies with lower metallicities. Dedicated observing campaigns with future facilities such as \emph{JWST}/NIRSpec will help shed light on the presence and extent of such a bias.

Another point worth raising is that, while both $Z_{\rm emiss}$ and $Z_{\rm abs}$ are quoted in terms of solar metallicity, each is traced by different elements ($Z_{\rm emiss}$ is traced by emission lines from various lighter metals (e.g. N, O) calibrated to oxygen abundances at low redshift; while $Z_{\rm abs}$ is a combination of heavier elements including Zn, Mn, and Fe -- see \citealt{Bolmer+19}). Thus, direct comparison of $Z_{\rm emiss}$ and $Z_{\rm abs}$ in this way assumes a uniform scaling of elemental abundances between solar abundances and those measured in this high-redshift galaxy. In reality, it is expected that abundance patterns will in fact vary from solar, although the relationship is complex \citep[e.g.][]{Nicholls+17}. While some high-redshift studies do favour an evolution in Fe/O from solar \citep[e.g.][]{Strom+17, Sanders+20}, more detailed studies are required to better constrain this relation across the full abundance range.

In this study, we have focused on absorption systems illuminated by GRBs.
Quasars are also well-studied sources of ionising radiation from which absorption metallicities may be computed. Using similar postprocessing methods to the one presented in this study, a $Z_{\rm abs}-Z_{\rm emiss}$ relation for quasar-illuminated galaxies could also be computed.
However, quasars will often outshine their host galaxies, increasing the challenges associated with measuring $Z_{\rm emiss}$ for the host galaxies of quasar-illuminated DLAs, making the predictions of such a model difficult to test. 
Because GRB afterglows fade relatively quickly, GRB host galaxies provide a unique opportunity to assess the difference between metallicity measurements obtained using emission-line spectroscopy and the DLA method. 
%Follow-up on the host galaxies of GRB120317A and GRB120909A using JWST's NIRSpec instrument would allow us to simultaneously explore the strength of the GRB metallicity bias function and the internal metallicity distributions of galaxies in the IllustrisTNG simulation. Once a large enough sample has been collected, this analysis could then be repeated using other simulations with different chemical enrichment prescriptions as an independent test of these sub-resolution models.

A recent analysis of metallicity gradients of galaxies in the TNG50 simulation has shown that at high redshifts, galaxies in the TNG50 simulation have steeper metallicity gradients than those observed \citep{Hemler+20}. This tension between the model and the data may be due to observational systematics associated with metallicity gradient measurements in high-redshift hosts, such as the flattening of observed metallicity gradients due to limitations in angular resolution or SNR of these observational studies \citep{Acharyya+20}, sensitivities to the range of radii over which the metallicity gradient is measured, or sample selection biases in observational studies; or it could be due to inaccuracies in the sub-resolution prescriptions of physical phenomena in the IllustrisTNG simulation suite, such as turbulence or stellar winds. Because neither measurement of $Z_{\rm abs}$ nor $Z_{\rm emiss}$ are affected by the angular resolution limitations associated with gathering IFU data of high-redshift galaxies, this method offers an independent probe of the dynamics of the ISM in the IllustrisTNG simulation suite. If, after observation of the two target GRB hosts with \emph{JWST} we were to discover that none of the theoretical $Z_{\rm abs}-Z_{\rm emiss}$ curves agree with our data, we may conclude that either the sub-resolution models of the IllustrisTNG simulation need to be refined further; or our model of GRB formation is not correct. We note that changing the internal metallicity distribution of these simulated galaxies will alter the predicted $Z_{\rm abs}-Z_{\rm emiss}$ relation of GRB host galaxies in a way that is difficult to quantify.% without repeating this analysis using other existing simulations with different chemical enrichment prescriptions.

Of the 32 GRB host galaxies for which $Z_{\rm abs}$ is known, 15 do not have meaningful brightness limits. 
A campaign to follow up these galaxies with 10m-class ground-based telescopes could increase the number of candidates for spectroscopy using \emph{JWST}'s NIRSpec instrument. Assuming that the sample of hosts without magnitude limits does not contain objects with unpublished non-detections, then it is intriguing to consider that 5-7 of the 17 galaxies with current detection (or published limits) are likely bright enough for \emph{JWST} follow up. Thus, the current sample of hosts with absorption metallicity could likely include $N \gtrsim 10$ objects where \emph{JWST} spectroscopy can measure emission metallicity. 
This would be sufficient to provide a well-constrained scaling relation that could be used as a tool to test chemical enrichment schemes in cosmological simulations on a sub-kiloparsec scale. 
Possibly, such a sample size could also provide constraints on the GRB bias function. In addition to \emph{JWST}, future progress will also be possible thanks to next-generation adaptive optics instrumentation for the Very Large Telescope currently being developed (the MAVIS imager and spectrograph\footnote{http://mavis-ao.org/mavis}).

\section{Conclusions} \label{sec:conc}

% The problem; one last time. (Summarise the intro)
Previous work \citep{Paper1} has shown that absorption metallicities computed from observations of GRB afterglows ($Z_{\rm abs}$) and gas-phase metallicities of galaxies determined using emission-line spectroscopy ($Z_{\rm emiss}$) exhibit systematic differences in the TNG100 simulation. In this paper, we explored whether the relation between the two quantities encodes information useful to both constrain GRB formation pathways, and inform models of chemical inhomogeneities in high redshift galaxies.

Using the IllustrisTNG simulation, theoretical models of the $Z_{\rm abs}-Z_{\rm emiss}$ relation for high-redshift GRB host galaxies are produced using a range of assumed maximum metallicities for GRB progenitors. We showed that different models of GRB formation predict different values of $Z_{\rm abs}$ for galaxies with fixed $Z_{\rm emiss}$; but the median values of $Z_{\rm emiss}$ for galaxies with a known value of $Z_{\rm abs}$ was found to be independent of the GRB metallicity bias model used. Furthermore, we found that the same GRB host galaxy (and galaxies with the same global metallicity $Z_{\rm emiss}$ in general) can have vastly different values of $Z_{\rm abs}$, depending on (i) the location of the GRB within the galaxy, and (ii) the orientation of the line-of-sight between the galaxy and an observer. Relationships between $Z_{\rm abs}$ and $Z_{\rm emiss}$ showed no evolution with redshift within the range explored ($2<z<4$).

Currently, only one GRB host galaxy has observational determinations of both $Z_{\rm abs}$ and $Z_{\rm emiss}$. After re-computing the value of $Z_{\rm emiss}$ using state-of-the-art Bayesian methods and an extended set of emission-line diagnostics (which returns an estimate of $Z_{\rm emiss}$ consistent with previous estimates by \citet{Friis+15} and \citet{Kruhler+15}), we showed that this single data point agrees best with GRB metallicity bias functions for which $Z_{\rm max} \sim 0.31 Z_\odot$. This metallicity threshold is similar to GRB threshold metallicities found in other studies \citep[e.g.][]{Vergani+15, Graham&Fruchter2017}; however, this single data point only has enough statistical power to rule out models of GRB formation for which $Z_{\rm max} < 0.07 Z_\odot$ with $95\%$ confidence. Hence, more data is needed in order to further constrain the $Z_{\rm abs}-Z_{\rm emiss}$ relation of GRB host galaxies.
This could be achieved by spectroscopic follow-up of the host galaxies of GRB000926, GRB070802, GRB090323, and GRB120327A (and potentially two others) using the forthcoming \emph{James Webb Space Telescope}.

While these two additional GRB hosts are the only known targets for which $Z_{\rm abs}-Z_{\rm emiss}$ can be measured in a reasonable amount of time, there are opportunities to increase the sample size. Currently there are 15 GRB hosts with measured $Z_{\rm abs}$ for which magnitude limits have not been obtained. Systematic photometric follow-up of these host galaxies using ground-based telescopes could arguably yield another handful of candidates brighter than AB magnitude 25 (based on the $\sim 25\%$ fraction of such sources in the upper portion of Table~\ref{tab:mangitudes}). Furthermore, future space missions for GRB detection and/or follow-up such as the SVOM mission\footnote{http://www.svom.fr/en/portfolio/the-svom-mission/} and the SkyHopper CubeSat\footnote{https://skyhopper.space} would offer the opportunity to expand the use of the method proposed here to shed new light on chemical enrichment processes and GRB formation in the first 1-2 billion years after the Big Bang.

\section*{Acknowledgements}

We would like to thank Dr. Antonino Cucchiara for insightful advice in interpreting absorption-line metallicity measurements and for useful discussions; the IllustrisTNG team for making their data public and providing lots of documentation and support; and the anonymous referee, whose feedback helped improve and shape this paper. This research was supported by the Australian Research Council Centre of Excellence for All Sky Astrophysics in 3 Dimensions (ASTRO 3D), through project number CE170100013. BM and AJC acknowledge support from Australian Government Research Training Program (RTP) Scholarships.

\section*{Data Availability}

This paper uses data products from the IllustrisTNG simulation suite \citep{TNG1, TNG2, TNG3, TNG4, TNG5} which are available for public download at \url{https://www.tng-project.org/data/}. Data used to construct any of the figures in the text are available from the corresponding author (BM) on reasonable request.

%%%%%%%%%%%%%%%%%%%%%%%%%%%%%%%%%%%%%%%%%%%%%%%%%%



\bibliographystyle{mnras}
\bibliography{biblio} % if your bibtex file is called example.bib

%\begin{thebibliography}{99}
%\end{thebibliography}

%%%%%%%%%%%%%%%%%%%%%%%%%%%%%%%%%%%%%%%%%%%%%%%%%%

%%%%%%%%%%%%%%%%% APPENDICES %%%%%%%%%%%%%%%%%%%%%
\newpage
\appendix

\section{Measuring gas-phase metallicity at high-redshift with emission lines}
\label{ap:gas_metallicity}

\subsection{Overview: The Challenges In Establishing a Universal Gas-Phase Abundance Scale}
\label{ap:gas_metal_overview}

Numerous methods exist for determining the metallicity of H{\sc ii} regions in galaxies from emission line spectroscopy. However different approaches do not always agree and reconciling these into a reliable universal abundance scale has proven challenging (see \citealt{MaiolinoMannucci19} and \citealt{Kewley+19} for recent reviews). 

Emission line strengths are dependent on electron temperature ($T_e$) in addition to abundance. Thus the preferred approach for establishing an absolute abundance scale is the so-called ``direct method’’ which requires the detection of a weak ``auroral’’ emission line (refer to \citealt{PerezMontero17} for a tutorial). The electron temperature can be explicitly determined from the ratio of this auroral line to a corresponding strong nebular line from the same ionic species (commonly [O {\sc iii}] $\lambda$4363 and [O {\sc iii}] $\lambda$5007), allowing for ionic abundances to be directly determined from emission line fluxes (although see \citealt{Nicholls+20} for the limitations of the direct method).

A key issue with the direct method is the requirement for the detection of these weak auroral lines which can be $\sim$100 times fainter than the strong nebular lines. Indeed, measurement of these auroral lines is challenging even in the local Universe, often requiring stacked spectra \citep[e.g.][]{AndrewsMartini13, Curti+17, Curti+20a}. While modest samples of direct method metallicities have been assembled at intermediate redshift \citep[$z\sim0.7$;][]{Jones+15, Ly+16}, only a handful of auroral line detections have been reported in the literature beyond $z\gtrsim1$ \citep{Patricio+18, Sanders+20}.

Strong-line methods offer an alternative approach in which metallicities are derived empirically from ratios of the brightest emission lines, using calibrations derived either from direct-method observations \citep{PettiniPagel04, Curti+20a} or stellar population synthesis and photoionisation modelling \citep{KewleyDopita02, Dopita+16, Kewley+19}.
Strong-line methods mitigate the detection rate issue faced by the direct method by relying on only the brightest emission lines. Gas-phase metallicity studies at high-redshift are almost exclusively done with strong-line methods \citep[e.g.][]{Erb+06, Maiolino+08, Zahid+14, Sanders+20b}.

One issue with strong-line based metallicities is that they can be very sensitive to assumptions of the physical conditions such as the ionisation conditions and chemical abundance ratios.
Recent studies of high-redshift galaxies have shown that relations of the diagnostic emission line ratios of $z\sim2$ galaxies are systematically offset from those typically observed in local galaxies, especially the widely used N2-BPT diagram \citep{Kewley+13, Steidel+14, Strom+17}.
These offsets are typically interpreted as systematic evolution with redshift of some combination of the ionisation parameter, shape of the ionising spectrum, and the N/O abundance ratio in H{\sc ii} regions \citep[e.g.][]{Steidel+14, Jones+15, Strom+17, Bian+20, Topping+20a, Topping+20b}. Such evolution may cause systematic offsets in metallicities derived in high-redshift galaxies from strong-line methods that are calibrated from local Universe measurements.

An additional class of strong-line methods are those that simultaneously constrain metallicity and select other physical parameters (especially ionisation parameter) from a set of observed emission line fluxes (e.g. IZI, \citealt{Blanc+15}; NebulaBayes, \citealt{Thomas+18}). These are less widely used, but automatically allow for evolution in any physical conditions considered in those models.

\subsection{Deriving Gas-Phase Metallicity For GRB121024A}
\label{ap:gas_metal_121024a}

\begin{figure}
\centering
\includegraphics[width=\columnwidth]{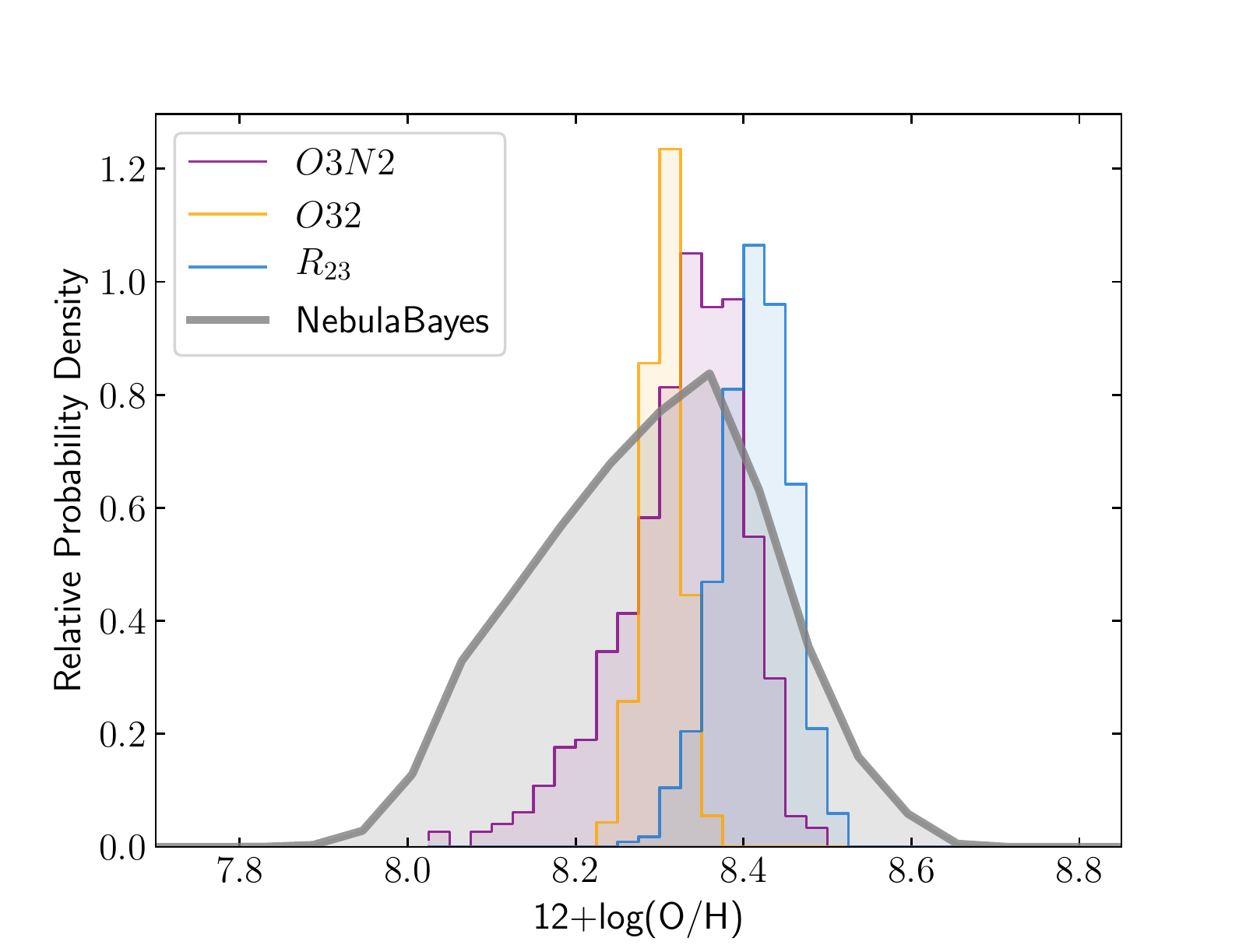}
\caption{Comparison of marginalised metallicity posterior obtained from NebulaBayes with metallicities obtained from the $O3N2$, $O32$ and $R_{23}$ strong-line diagnostics using calibrations provided by \citet{Bian+18} based on local analogues of high-redshift galaxies. The histogram shown for each diagnostic is formed using the bootstrapping method described in Appendix~\ref{ap:gas_metal_121024a} and then are re-normalised arbitrarily for visual clarity. Normalisation of the NebulaBayes posterior is also arbitrary.}
\label{fig:ap_logOH_hist}
\end{figure}

\begin{figure}
\centering
\includegraphics[width=\columnwidth]{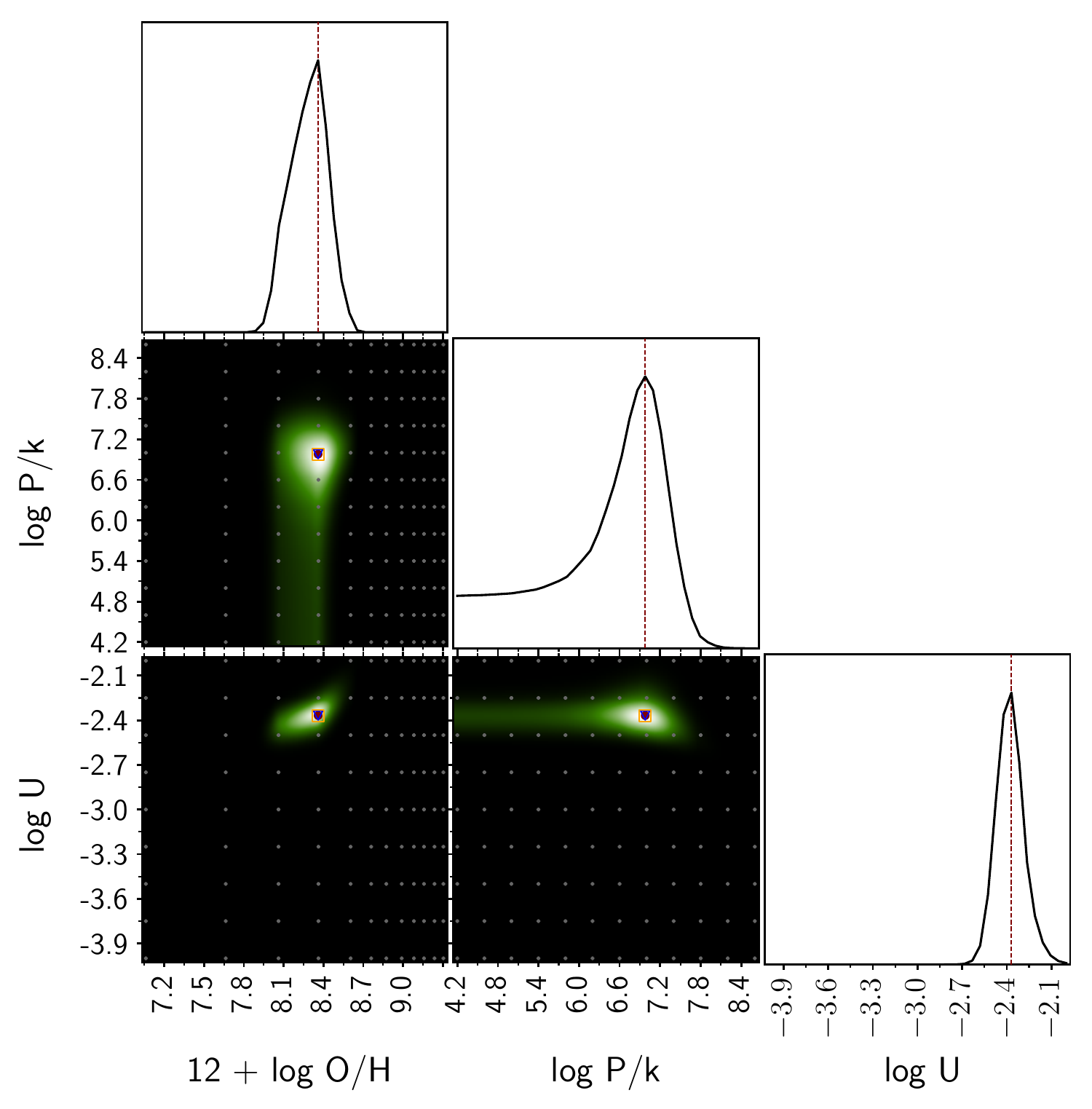}
\caption{Marginalised probability density functions of metallicity (12+log($O/H$)), ionisation parameter (log($U$)), and ISM pressure (log($P/k$)) from NebulaBayes \citep{Thomas+18}.
Vertical dashed lines show peak of 1D marginalised PDFs. In the panels showing 2D marginalised PDFs, the red filled circle shows the model defined by peaks of 1D PDFs, the blue triangle shows the location of the pead in the 2D marginalised PDF, and the yellow square shows the projected peak of the full nD PDF. In the case of the host galaxy of GRB121024A, these peaks lie in the same location.
}
\label{fig:ap_NB_post}
\end{figure}

Emission line spectroscopy of the GRB121024A host galaxy was first reported by \citet{Friis+15} and then again by \citet{Kruhler+15} with flux measurements obtained for the following strong-lines: [O {\sc ii}] $\lambda\lambda$3726, 29, H$\beta$, [O {\sc iii}] $\lambda\lambda$4959, 5007, H$\alpha$, and [N {\sc ii}] $\lambda$6583. The absence of constraints on the flux of the [O {\sc iii}] $\lambda$4363 auroral emission line precludes the application of the direct method and thus the gas-phase metallicity must be inferred from strong-line methods.

\citet{Kruhler+15} derive a gas-phase metallicity of $12+\text{log}(O/H)=8.41^{+0.11}_{-0.12}$ using a multi-diagnostics method. In this approach several metallicity probability distributions are obtained for a number of available diagnostic strong-line ratios using the calibrations of \citet{Nagao+06} and \citet{Maiolino+08}. The final metallicity PDF is obtained by combining these individual diagnostic-by-diagnostic distributions.
The diagnostic line ratios used are $N2$, $O3N2$, $N2O2$, and $R23$, which are defined below:

\begin{gather}
    \label{eq:line_ratio_N2O2}
    N2O2 = \text{log}_{10}\left(\text{[N {\sc ii}]/[O {\sc ii}]}\right)\\
    \label{eq:line_ratio_O32}
    O32 = \text{log}_{10}\left(\text{[O {\sc iii}]/[O {\sc ii}]}\right)\\
    \label{eq:line_ratio_R23}
    R_{23} = \text{log}_{10}\left(\frac{\text{[O {\sc iii}]}\lambda4959 + \text{[O {\sc iii}]}\lambda5007 + \text{[O {\sc ii}]}}{H\beta}\right)\\
    \label{eq:line_ratio_N2}
    N2 = \text{log}_{10}\left(\text{[N {\sc ii}]/H}\alpha\right)\\
    \label{eq:line_ratio_R3}
    R3 = \text{log}_{10}\left(\text{[O {\sc iii}]/H}\beta\right)\\
    \label{eq:line_ratio_O3N2}
    O3N2 = R3 - N2
\end{gather}
\smallskip

where [N {\sc ii}] = [N {\sc ii}] $\lambda$6583, [O {\sc ii}] = ([O {\sc ii}] $\lambda$3726 + [O {\sc ii}] $\lambda$3729), and  [O {\sc iii}] = [O {\sc iii}] $\lambda$5007 unless otherwise specified.

One issue with the above approach is its reliance on calibrations obtained from local Universe measurements.
As highlighted in Appendix~\ref{ap:gas_metal_overview}, high-redshift galaxies have been observed to inhabit a locus on the N2-BPT diagram offset from that observed for local galaxies, suggesting some degree of redshift evolution in the physical conditions in H{\sc ii} regions.
A number of explanations for this offset have been proposed including evolution in the N/O abundance ratio, ionisation parameter, or shape of the ionising spectrum \citep[e.g.][]{Steidel+14, Jones+15, Strom+17, Bian+20, Topping+20a, Topping+20b}.
The evolution of diagnostic emission line ratios with redshift suggests that large systematic uncertainties may be introduced by deriving metallicities with strong-line methods that have been calibrated for typical local galaxies, especially those using the $N2$ ratio.

To address this issue, \citet{Bian+18} provide strong-line calibrations based on a subsample of SDSS galaxies selected based on their position on the N2-BPT diagram to represent analogues of high-redshift galaxies.
\citet{Sanders+20} found that these calibrations closely matched direct method metallicities from stacked spectra of $z\sim2$ galaxies.
The \citet{Bian+16} sample, on which the \citet{Bian+18} calibration is based, uses position on the N2-BPT diagram as its primary selection criterion. The sample selection is based on the locus inhabited by the $z\sim2$ sample from \citet{Steidel+14}, although a slightly stricter lower cutoff is imposed to reduce the contaminating impact of ``normal'' local galaxies.
With measured BPT line ratios of $N2 = -1.04 \pm 0.20$ and $R3 = 0.57 \pm 0.04$, we find that the host galaxy of GRB121024A is indeed offset from the local N2-BPT sequence and falls within the region inhabited by the \citet{Steidel+14} sample, albeit at the lower edge, outside the \citet{Bian+16} selection.

An alternative approach to accounting for possible evolution in the physical conditions of the emitting H{\sc ii} regions is to simultaneously fit for these other physical parameters when determining the metallicity from emission line ratios.
This can be achieved by a number of publicly available tools which fit sets of emission line fluxes to photoionisation model predictions. Examples of this include \texttt{NebulaBayes} \citep{Thomas+18} which fits for ionisation parameter and ISM pressure alongside metallicity, and {\sc HII-CHI-MISTRY} \citep{PerezMontero14} which can be be used to compute ionisation parameter and nitrogen-to-oxygen abundance ratio in addition to metallicity.

We re-derive the metallicity of the host galaxy of GRB121024A using \texttt{NebulaBayes} as well as the updated high-redshift calibrations from \citet{Bian+18} using the emission line flux measurements reported in \citet{Kruhler+15}.
This is so as to attempt to better allow for evolution in physical conditions, including ionisation parameter, rather than relying on strong-line diagnostics calibrated from local samples.

Using the \citet{Bian+18} calibrations, we derive metallicities from $O3N2$ and $R_{23}$ diagnostics (as in \citealt{Kruhler+15}) as well as $O32$, which \citet{Sanders+20} cite as the best diagnostic for $z\sim2$ galaxies up to $Z \lesssim 1/3 \cdot Z_\odot$ on the basis of its clear monotonic trend with metallicity.
We use a bootstrapping method in which we draw 1000 instances of each emission line from a normal distribution centred on the reported emission line flux with a standard deviation of the reported 1-$\sigma$ measurement uncertainty.
The values we obtain are $Z_{O3N2} = 8.34^{+0.06}_{-0.08}$, $Z_{R23} = 8.42^{+0.04}_{-0.05}$, and $Z_{O32} = 8.31^{+0.02}_{-0.02}$, where $Z=12+\text{log}(O/H)$. The quoted uncertainties reflect the central 68\% interval of the resulting histogram. Re-normalised histograms obtained from this analysis are shown in Figure~\ref{fig:ap_logOH_hist}.\footnote{It should be noted that the metallicity probability distribution obtained with $R_{23}$ here partially extends beyond the valid metallicity calibration range from \citet{Bian+18}.}

The dominant source of uncertainty in reality is the systematic uncertainty. The extent that these values disagree is in some sense a lower bound on the systematic uncertainty. Rather than combine the metallicity probably distributions obtained from each diagnostic, we instead consider model fitting from the NebulaBayes package \citep{Thomas+18}. This code compares all observed emission line fluxes to model grids to simultaneously infer metallicity, ionisation parameter and ISM pressure. Parameter estimates from the marginalised metallicity, ionisation parameter and ISM pressure probability density functions using the \citet{Kruhler+15} emission line fluxes are as follows: $12+\text{log}(O/H)=8.36^{+0.12}_{-0.24}$, $\text{log}(U)=-2.37^{+0.11}_{-0.11}$, $\text{log}(P/k)=6.98^{+0.23}_{-1.74}$ where quoted uncertainties show the 68\% confidence interval. The marginalised posterior distributions output by NebulaBayes are shown in Figure~\ref{fig:ap_NB_post}. By considering the ionisation parameter, this approach better allows for systematic offsets observed at high-redshifts than fixed calibrations. Indeed, the ionisation parameter measured here is consistent with values observed for $z\sim2$ galaxies \citep{Strom+18}.

Since the 68\% confidence interval on the metallicity parameter estimate obtained from NebulaBayes encompasses a metallicity range that includes the value obtained by each of the three \citet{Bian+18} strong-line diagnostics (see Figure~\ref{fig:ap_logOH_hist}), we adopt this as the gas-phase metallicity of the host galaxy of GRB121024A.

The NebulaBayes value for the metallicity ($12+\text{log}(O/H)=8.36^{+0.12}_{-0.24}$) is 0.05 dex lower than the value obtained by \citet{Kruhler+15} ($12+\text{log}(O/H)=8.41^{+0.11}_{-0.12}$). Relations derived by \citet{Bian+18} indicate that at values of log([N {\sc ii}]/H$\alpha) \gtrsim -1.2$, their local analogues of high-redshift galaxies have lower metallicity than local reference galaxies at fixed N2. The host galaxy of GRB121024A has log([N {\sc ii}]/H$\alpha) = -0.992$, suggesting that metallicity diagnostics calibrated from typical local galaxies (as used in \citealt{Kruhler+15}) may overestimate metallicities when derived using the [N {\sc ii}] emission line. Given that all diagnostics are given equal weighting in the multi-diagnostic approach used by \citet{Kruhler+15}, this effect could help to explain the difference in measured metallicity.

% \section{Algorithm for computing $Z_{\rm abs}$ }

% Can insert content from thesis appendix here if you want.
%%%%%%%%%%%%%%%%%%%%%%%%%%%%%%%%%%%%%%%%%%%%%%%%%%

% Don't change these lines
\bsp	% typesetting comment
\label{lastpage}
\end{document}